\documentclass[reqno]{amsart}

\usepackage[unicode,psdextra]{hyperref}
\hypersetup{
    colorlinks,
    linkcolor={red!50!black},
    citecolor={blue!50!black},
    urlcolor={blue!90!black}
}

\usepackage{graphicx}
\usepackage{xcolor}
\usepackage{xfrac}
\usepackage{enumitem}

\usepackage[english]{babel}
\usepackage[T1]{fontenc}
\usepackage{libertine}

\newcommand{\pd}{\partial}

\newcommand{\dd}{\mathrm{d}}
\newcommand{\ii}{\mathrm{i}}
\newcommand{\pdv}[2]{\frac{\partial #1}{\partial #2}}
\newcommand{\dv}[2]{\frac{\mathrm{d} #1}{\mathrm{d} #2}}
\DeclareMathOperator*{\Res}{Res}

\renewcommand{\Re}{\,\mathrm{Re}}

\makeatletter
\renewcommand\subsubsection{\@startsection{subsubsection}{3}%
  \z@{.5\linespacing\@plus.7\linespacing}{-.5em}%
  {\normalfont\bfseries}}
\makeatother

\makeatletter
\renewcommand\paragraph{\@startsection{paragraph}{4}%
  \z@{.5\linespacing\@plus.7\linespacing}{-.5em}%
  {\normalfont\bfseries}*}
\makeatother

\usepackage{csquotes}  
\usepackage[
    backend=biber,
    style=phys,
    citestyle=numeric-comp,
    sorting=none,
    eprint=true,
    doi=false,
    url=false,
    biblabel=brackets
]{biblatex}

\begin{document}

\title{A first passage problem for a Poisson counting process with a linear moving boundary}

\author[Ivan Burenev, 
        Michael J. Kearney, 
        and Satya N. Majumdar]
       {Ivan N. Burenev$^{1,\dagger}$, 
       Michael J. Kearney$^{2}$, 
       and Satya N. Majumdar$^{1}$}

\thanks{$^1$LPTMS, CNRS, Universit\'e Paris-Saclay, 91405 Orsay, France}
\thanks{$^2$University of Surrey, Guildford, Surrey, GU2 7XH, United Kingdom}
\thanks{$^\dagger$\textit{E-mail address:} inburenev@gmail.com}

\begin{abstract}
    The time to first crossing for the Poisson counting process with respect to a linear moving barrier with offset is a classic problem, although key results remain scattered across the literature and their equivalence is often unclear. 
    Here we present a unified and pedagogical treatment of two approaches: the direct time-domain approach based on path-decomposition techniques and the Laplace-domain approach based on the Pollaczek-Spitzer formula. 
    Beyond streamlining existing derivations and establishing their consistency, we leverage the complementary nature of the two methods to obtain new exact analytical results.
    Specifically, we derive an explicit large deviation function for the first-passage time distribution in the subcritical regime and closed-form expressions for the conditional mean first-passage time for arbitrary offset.
    Despite its simplicity, this first crossing process exhibits non-trivial critical behavior and provides a rare example where all the main results of interest can be derived exactly.
\end{abstract}

\maketitle

\vspace{-.6cm}

\tableofcontents

\clearpage

\section{Introduction}\label{sec:introduction}

\par 
In this paper we consider a Poisson process, i.e., a random counting process $N(t)$ that tracks discrete events occurring in continuous time. The fundamental assumption is that events happen independently with a constant rate $r$. This means that in any infinitesimal time interval $\dd t$, the probability of observing an event is $r \dd t$, and this probability does not depend on the history of the process. Starting from this simple definition, one can readily derive that the number of events occurring in the time interval $(0, t)$ follows a Poisson distribution:
\begin{equation}\label{eq:poisson_dist}
    \mathrm{Pr} \big[ N(t) = n \big] = \frac{(rt)^n}{n!} e^{-rt}.
\end{equation}
Equivalently, one can define the process by specifying that the waiting times between consecutive events are independent and exponentially distributed random variables with mean $1/r$; each time an event occurs, $N(t)$ increases by one.  By conditioning on the time $t'$ at which the $n$-th (last) jump occurs, we can decompose the probability into three independent contributions: (i) exactly $n-1$ jumps occur in the time interval $[0, t']$, with probability $\mathrm{Pr} \big[ N(t') = n-1\big]$; (ii) one jump occurs in the infinitesimal interval $[t', t'+\dd t']$, with probability $r \, \dd t'$; (iii) no further jumps occur in the remaining interval $[t'+\dd t', t]$, with probability $e^{-r (t-t')}$. Integrating over all possible values of $t'$, we obtain the recurrence relation
\begin{equation}\label{eq:recurrence_Poisson}
    \mathrm{Pr} \big[ N(t) = n\big] = \int_{0}^{t} \dd t'\; \mathrm{Pr} \big[ N(t') = n-1\big] \, r \, e^{-r (t-t')},
\end{equation}
with the initial condition $\mathrm{Pr} \big[ N(t) = 0\big] = e^{-r t}$ (the probability of observing no jumps). One can solve this recurrence relation explicitly and verify that it yields the Poisson distribution \eqref{eq:poisson_dist}.

\par Throughout this work, we set $r = 1$ by choosing our time units appropriately. With this scaling, the distribution \eqref{eq:poisson_dist} becomes
\begin{equation}\label{eq:poisson_dist_unit}
    \mathrm{Pr} \big[ N(t) = n\big] = \frac{t^n}{n!} e^{-t},
\end{equation}
and the process accumulates on average one event per unit time.

\par
\begin{figure}[ht]
\centering
\includegraphics{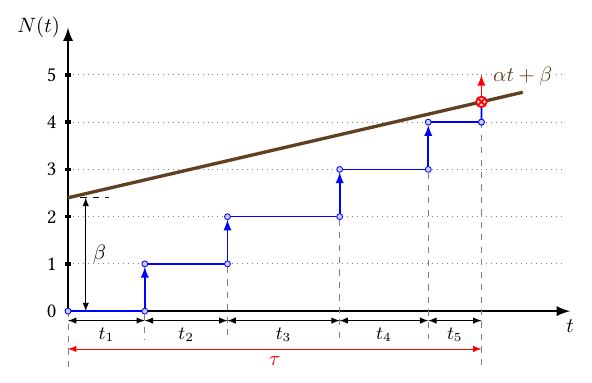}
\caption{Schematic of a sample path of the Poisson process showing its evolution in relation to the linear moving boundary with positive offset $B(t) = \alpha t + \beta$. In this example, the path crosses the boundary for the first time on the fifth jump at time $\tau$. The time intervals $t_i$ between jumps are independent random variables drawn from the exponential distribution with unit rate.}\label{fig:trajectory}
\end{figure}

 Our aim is to study the first-passage properties of this process with respect to a linear moving boundary having slope $\alpha\ge0$ and initial offset $\beta\ge0$. More precisely, we are interested in the statistics of the first-passage time defined as
\begin{equation}\label{eq:tau=definition}
    \tau \equiv \min\{t : N(t) > \alpha t + \beta \}.
\end{equation}
This quantity represents the first time at which the counting process crosses the moving boundary (see Fig.~\ref{fig:trajectory}). Clearly, $\tau$ depends on the realization of the process and hence is a random variable.

\par
\begin{figure}[ht]
\centering
\includegraphics{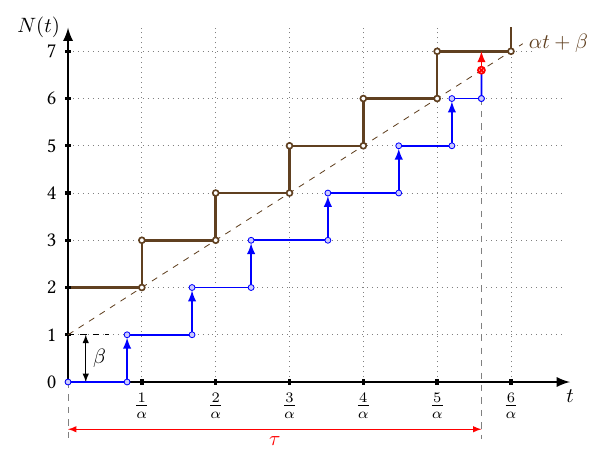}
\caption{
Schematic representation of the D/M/1 queue with $k=2$ initial 
customers. The Poisson process (blue) counts the number of customers served by time $t$. The piecewise constant function (brown) represents the total number of customers who have arrived by time $t$. The linear boundary $B(t) = \alpha t + \beta$ with $\beta=k-1$ serves as the lower enveloping curve for this piecewise constant function. The boundary crossing corresponds to the moment when the number of served customers equals the number of customers who have arrived, hence the queue is depleted. }\label{fig:trajectory_queue}
\end{figure}

 The model we consider admits several interpretations.  One such comes from queueing theory (see \cite{Bhat2015} for a pedagogical introduction). Consider a queue that initially has $k\ge1$ customers waiting and suppose that (i) new customers arrive at deterministic time intervals $1/\alpha$, $2/\alpha$, $3/\alpha$, $\ldots$ and join the queue; (ii) the service times are independent random variables following an exponential distribution with unit rate. 
In queueing theory, this is known as a D/M/1 queue (deterministic arrivals, Markovian service times, single-server queue). The Poisson process $N(t)$ counts the number of customers served by time $t$, while the total number of customers who have entered the system by time $t$ is $k + \lfloor \alpha t \rfloor$. Since $N(t)$ can only cross the boundary at service completion times (when $N(t)$ jumps), the first-passage time $\tau$ corresponds to the busy period, i.e., the time until the queue becomes empty (see Fig.~\ref{fig:trajectory_queue}).

\par 
Another interpretation comes from predator-prey models. The process $N(t)$
can be viewed as a predator making unit-length jumps with 
exponentially distributed waiting times while pursuing a prey that moves away 
deterministically with constant velocity $\alpha$. The first-passage time 
then corresponds to the capture time. Related problems involving 
stochastic pursuit and moving boundaries are discussed in e.g. \cites{KR-96,KR-96b,RK-99,OVKK-09}.

\par First-passage problems, which in a broader sense concern the time at which a stochastic process reaches a boundary, are longstanding and very easy to formulate. Unfortunately, they are also known to be difficult to solve analytically, even for simple one-dimensional systems, except in special cases (see e.g. \cite{BMS-13} for a physical perspective and \cite{AS-15} for a more mathematical point of view; see also \cites{Redner2001,MajumdarSchehr2024} for comprehensive monographs). The problem of the first-passage time of the Poisson process with respect to a linear moving boundary is one of these rare tractable cases, and consequently it has received considerable attention in the literature. 

\par Let us briefly overview the main known results, highlighting two seminal papers. The first is \cite{BD-57}, where the double Laplace transform of the survival probability (the probability that the crossing has not occurred up to time $t$) was obtained in closed form (see \cite{T-70} for an alternative derivation). The derivation was based on the Spitzer combinatorial identity \cite{S-56} (which was also obtained independently by Pollaczek in \cite{P-52}). The second is \cite{P-59}, in which this probability was computed directly in the time domain using path decomposition techniques. Similar techniques were later employed in \cite{Z-91,DP-97} and extended to more general boundary problems in \cite{T-65,G-66,G-93,Lehmann1998,PSZ-99} (see \cite{Z-17} for a review).

\par Given this long history with numerous contributions, one can erroneously conclude there is nothing more to say. However, at a fundamental level we note that the Laplace transform obtained in \cite{BD-57} is challenging to invert analytically. Demonstrating that it yields results equivalent to those of \cite{P-59} is nontrivial; to the best of our knowledge, this connection has not been established explicitly in the literature. Moreover, largely due to the difficulty of inverting the Laplace transform, much of the subsequent work has focused on direct time-domain calculations, leaving the Laplace transform approach relatively obscure.

\par The purpose of this paper is thus twofold. The first goal is mostly pedagogical: we present side by side both approaches~--- the direct time-domain calculation and the Laplace transform derivation. For the latter approach, instead of following \cite{BD-57}, we adapt the method of \cite{BM-25}, computing the double Laplace transform via a different route that we find more transparent, even though it also relies on the Spitzer result \cite{S-56}.  We believe this unified presentation will make the powerful Laplace transform techniques more accessible to the physics community. Second, we analyze the Laplace transform to derive asymptotic properties of the first-passage time distribution and obtain closed-form expressions for the mean first-passage time. These results appear to be new and provide additional insights into the problem.

\par The rest of the paper is organized as follows. In Section~\ref{sec:ModelAndResults} we give a more detailed overview of the known results and state the original contributions of the paper.  
In Section~\ref{sec:TwoFrameworks}, we develop both the time- and Laplace-domain approaches, obtaining the time-domain representation  and its Laplace transform. 
Section~\ref{sec:Extreme cases} treats the special cases $\beta=0$ and $\beta\to\infty$. The former is simple enough to rigorously establish the equivalence of the two approaches while building the intuition needed for the general offset $\beta$, without excessive technical details. The latter highlights the power of the Laplace-domain approach for asymptotic analysis. In Section~\ref{sec:general_offset}, we tackle the general offset, showing how the two approaches complement each other to yield closed-form expressions for the mean first-passage time. Finally, we conclude in Section~\ref{sec:Conclusion}. 

\par 
Technical details are deferred to three appendices. Appendix~\ref{sec:app_LambertW} summarizes essential properties of the Lambert $W$-function, a special function that appears throughout our analysis. Appendix~\ref{sec:app_Q_Laplace} provides the detailed derivation of the explicit Laplace transform representation. Appendix~\ref{sec:app_Sinfty} verifies the consistency of the two approaches by rederiving certain asymptotic results from the time-domain perspective.

\section{The model and the main results}\label{sec:ModelAndResults}

\par In this section, we provide a detailed overview of the known results for the first-passage problem of the Poisson process with respect to a linearly moving boundary, and then present the original contributions of this work.

\subsection{Historical background} 
\par The first important result in this area dates back to 1957, when Baxter and Donsker \cite{BD-57}, relying on the Spitzer combinatorial identity \cite{S-56}, obtained a closed-form representation for the double Laplace transform of the survival probability for a broad class of processes (see also \cite{T-70} for an alternative derivation). The problem of a Poisson process crossing a linear boundary was then addressed as a special case. Let us state this result more precisely.

\par Denote by $S(\tau\,\vert\,\beta)$ the survival probability, i.e., the probability that the crossing has not occurred up to time $\tau$ given the initial offset $\beta$, and by $\hat{S}(\rho\,\vert\,\lambda)$ its double Laplace transform
\begin{equation}\label{eq:hatS(rho|lambda)=def_res}
    \hat{S}(\rho\,\vert\,\lambda) \equiv 
    \int_0^{\infty} \dd \beta\, e^{-\lambda\beta}
    \int_{0}^{\infty} \dd \tau \, e^{-\rho \tau} 
    S(\tau\,\vert\,\beta).
\end{equation}
Then, in the notation of the present paper and after some algebraic manipulations, the result of \cite{BD-57} can be written as
\begin{equation}\label{eq:hat(S)=explicit_res}
    \hat{S}(\rho\,\vert\,\lambda)= 
    \frac{1}{\lambda}
    \frac{1}{\left(\rho + 1 - \alpha\lambda - e^{-\lambda}\right) }
    \left(1 - \frac{\lambda}{\frac{\rho+1}{\alpha} + W_0\left[ 
    -\frac{1}{\alpha} e^{-{\frac{\rho+1}{\alpha}}} 
    \right]} \right),
\end{equation}
where $W_0$ is the principal branch of the Lambert $W$-function \cite{CGHJK-96}, a special function defined as a solution of $W(z) e^{W(z)} = z$  (see Appendix~\ref{sec:app_LambertW} for the details).   
\par Another natural quantity is the probability density $\mathbb{P}[\,\tau\,\vert\,\beta\,]$ of the first-passage time $\tau$, which is related to the survival probability via
\begin{equation}\label{eq:S(t|beta)=def}
    S(\tau\,\vert\,\beta) = 
        1 - \int_{0}^{\tau} \dd \bar{\tau}\,\mathbb{P}\left[\,\bar{\tau}\,\vert\,\beta\,\right],
    \qquad
    \mathbb{P}\left[\,\tau\,\vert\,\beta\,\right] 
        = - \pdv{}{\tau} \Big[ S(\tau\,\vert\,\beta) \Big].
\end{equation} 
From \eqref{eq:hat(S)=explicit_res} and \eqref{eq:S(t|beta)=def} we can find the Laplace transform of the first-passage probability density
\begin{equation}\label{eq:hat(Q)=def}
    \hat{Q}(\rho\,\vert\,\lambda) \equiv
    \int_{0}^{\infty} \dd \beta\,  e^{-\lambda\beta}
    \int_{0}^{\infty} \dd \tau\,  e^{-\rho\tau} 
    \, 
    \mathbb{P}[\tau\,\vert\,\beta].
\end{equation}
Indeed, due to \eqref{eq:S(t|beta)=def}, integrating by parts yields
\begin{equation}
    \hat{Q}(\rho\,\vert\,\lambda) = -
    \int_{0}^{\infty} \dd \beta\,  e^{-\lambda\beta}
    \left\{
    -1
    +\rho \int_{0}^{\infty} \dd \tau\,  e^{-\rho\tau} 
    \, 
    S(\tau\,\vert\,\beta)
    \right\}.
\end{equation}
Computing the integral with respect to $\beta$ in the first term and recognizing $\hat{S}(\rho\,\vert\,\beta)$ in the second we obtain
\begin{equation}\label{eq:hatS=1-hatQ}
    \hat{S}(\rho\,\vert\,\lambda) = \frac{1}{\rho}\left(\frac{1}{\lambda} - \hat{Q}(\rho\,\vert\,\lambda)\right).
\end{equation}

\par In principle, the survival probability, and hence the first-passage probability density, can be found from \eqref{eq:hat(S)=explicit_res} by inverting the double Laplace transform. This inversion is, however, highly nontrivial and has not been carried out. Instead, shortly after, in 1959, Pyke made significant progress \cite{P-59} by computing the survival probability $S(\tau\,\vert\,\beta)$, as well as the probability that the process never crosses the boundary $S_\infty(\beta)$, directly in the time domain using path-decomposition techniques. Specifically, he showed  that the survival probability is given by 
\begin{equation}\label{eq:S(t,beta)=result_time_domain_res}
    S(\tau\,\vert\,\beta) = \sum_{n=0}^{\lfloor\alpha \tau+\beta
    \rfloor} 
        \frac{(\alpha \tau + \beta - n)}{\alpha^n} 
        \sum_{j=0}^{\lfloor\beta\rfloor} 
        \frac{(j-\beta)^j(\alpha \tau + \beta - j)^{n-j-1}}
             {j!(n-j)!}e^{-\tau},
\end{equation}
where $\lfloor \ldots \rfloor $ is the floor function. Similarly, one can obtain the probability density of $\tau$ as
\begin{equation}\label{eq:P=result_time_domain_res}
\mathbb{P}\big[\,\tau\,\vert\,\beta\,\big] 
    = \frac{(\alpha \tau + \beta - \lfloor\alpha \tau + \beta
            \rfloor)}
           {\alpha^{\lfloor\alpha \tau+\beta\rfloor}} 
    \sum_{j=0}^{\lfloor\beta\rfloor} 
        \frac{(j-\beta)^j(\alpha \tau + \beta - j)^{\lfloor\alpha 
              \tau+\beta\rfloor-j-1}}
             {j!(\lfloor\alpha \tau+\beta\rfloor - j)!}e^{-\tau}.
\end{equation}
We note that although a direct consequence of the known result \eqref{eq:S(t,beta)=result_time_domain_res}, the explicit representation \eqref{eq:P=result_time_domain_res} for the probability distribution to the best of our knowledge has not been presented in the literature.

\par The final piece of the results obtained by the direct calculation in the time-domain is the closed-form expression for the probability that the process never crosses the boundary,
\begin{equation}\label{eq:Sinfty=def}
    S_{\infty}(\beta) \equiv \lim_{\tau\to\infty} S(\tau\,\vert\,\beta)=
        1 - \int_{0}^{\infty} \dd \bar{\tau}\,\mathbb{P}\big[\,\bar{\tau}\,\vert\,\beta\,\big] .
\end{equation}
It takes the form 
\begin{align}
\label{eq:Sinfty(beta)=a<1res}
    & \alpha \le 1:\qquad  S_\infty\left(\beta\right) = 0,\\
\label{eq:Sinfty(beta)=a>1res}
    & \alpha > 1:\qquad  S_\infty\left(\beta\right) = \left(1-\frac{1}{\alpha}\right)
        \sum_{j=0}^{ \lfloor \beta \rfloor } 
         \frac{1}{\alpha^j}\frac{(-1)^{j}}{j!}
        \left(\beta-j\right)^j 
        e^{\frac{\beta - j}{\alpha}}.
\end{align}
This results has a very simple physical interpretation. Since the boundary grows linearly with velocity $\alpha$ while the process grows on average with unit velocity, the parameter $\alpha$ plays a critical role: for $\alpha < 1$, the process eventually overtakes the boundary almost surely, while for $\alpha > 1$, there is a non-zero probability \eqref{eq:Sinfty(beta)=a>1res} that the process never reaches the boundary.

\par Remarkably, the equivalence between the Laplace-domain representation and the time-domain result is far from obvious. Computing the Laplace transform of \eqref{eq:S(t,beta)=result_time_domain_res} directly is obstructed by discontinuities from the floor function and summation limits that depend on the integration variables. Conversely, inverting \eqref{eq:hat(S)=explicit_res} via standard residue calculus techniques produces an infinite series that does not manifestly reduce to \eqref{eq:S(t,beta)=result_time_domain_res}. This nontrivial equivalence poses an interesting mathematical problem in its own right, and, to the best of our knowledge, has not been proved.

\par Despite their explicit nature, the time-domain representations \eqref{eq:S(t,beta)=result_time_domain_res} and \eqref{eq:P=result_time_domain_res} present significant computational challenges for several natural questions. For instance, extracting the large-time or large-offset behavior of the survival probability requires careful treatment of the summation limits and the discontinuous terms. 
Another quantity of interest is \emph{conditional moments} of the distribution, i.e., the moments of the first passage time computed for the trajectories in which the crossing has happened,
\begin{equation}\label{eq:Ec[t^k]=def}
    \mathbb{E}_c\left[\,\tau^k\,\vert\,\beta \, \right] \equiv 
    \frac{1}{1-S_\infty(\beta)} \int_{0}^{\infty}
    \dd \tau\, \tau^k\, \mathbb{P}\left[\,\tau\,\vert\,\beta\,\right].
\end{equation}
Hereafter, we will often omit the qualifier, and refer to \eqref{eq:Ec[t^k]=def} as simply the moments. 
Computing \eqref{eq:Ec[t^k]=def} from the direct time-domain representation \eqref{eq:P=result_time_domain_res} of the probability distribution becomes quite involved technically. In contrast, the Laplace-domain representation \eqref{eq:hat(S)=explicit_res} offers a natural framework for such calculations. As we demonstrate in this paper, systematic expansion and inversion of the Laplace transform yields closed-form results for moments and asymptotic properties that would be difficult to extract from the time-domain expressions alone.

\subsection{Main results}\label{sec:MainResults}
Below we present an overview of the results obtained in the present paper. They fall into three parts: the first concerns the asymptotic behavior of the survival probability at large times, the second addresses the large-offset limit, and the third provides a closed-form representation for the conditional mean first-passage time for the general offset $\beta$. 

\paragraph{Asymptotic behavior of the survival probability} 
\par We begin with the survival probability $S(\tau\,\vert\,\beta)$, which characterizes the fraction of trajectories that avoid crossing the boundary up to time $\tau$. We show that the survival probability decays to its asymptotic value $S_\infty(\beta)$ exponentially fast,
\begin{equation}\label{eq:S-Sinf ~ e^}
    \alpha\ne1: \qquad S(\tau\,\vert\,\beta) - S_{\infty}(\beta)
    \underset{\tau\to\infty}{\asymp} \exp\left[-\frac{\tau}{\xi(\alpha)}\right].
\end{equation}
The asymptotic value $S_\infty(\beta)$ is the probability that the process never crosses the boundary, and is given by \eqref{eq:Sinfty(beta)=a<1res} and \eqref{eq:Sinfty(beta)=a>1res}. The characteristic time scale reads
\begin{equation}\label{eq:xi=answer_res}
    \xi(\alpha) = \frac{1}{1-\alpha+\alpha\log\alpha}.
\end{equation}
We emphasize that $\xi(\alpha)$ does not depend on the offset $\beta$ and the functional form \eqref{eq:xi=answer_res} is the same on both sides of the critical point $\alpha=1$. A similar universal exponential decay was established for the dual process in \cite{B-25} in quite general settings (we discuss the notion of <<duality>> in Sec.~\ref{sec:LaplaceDomainApproach}).

\par As the slope approaches the critical value $\alpha \to 1$, the timescale \eqref{eq:xi=answer_res} diverges as $\xi(\alpha) \sim 2(\alpha - 1)^{-2}$, and the exponential decay in \eqref{eq:S-Sinf ~ e^} crosses over to a slower algebraic one,
\begin{equation}\label{eq:S(tau|beta)~critical_res}
    \alpha=1:\qquad
    S(\tau\,\vert\,\beta)
    \underset{\tau\to\infty}{\sim}
    \frac{1}{\sqrt{2\pi\tau}}
    \sum_{j=0}^{\lfloor \beta \rfloor}
    \frac{(-1)^{j}}{j!}(\beta-j)^j e^{\beta-j}.
\end{equation}
A direct consequence of this algebraic decay is the divergence of all conditional moments. Since the survival probability decays only algebraically as $\tau^{-\frac{1}{2}}$ for large $\tau$, the first-passage time distribution has a power-law tail $\tau^{-\frac{3}{2}}$ as $\tau\to\infty$, and the integrals \eqref{eq:Ec[t^k]=def} defining the conditional moments diverge. We find that the divergent behavior near the critical point is given by
\begin{equation}\label{eq:Ec=divergence_res}
    \mathbb{E}_{c}\left[\,\tau^{k}\,\vert\,\beta\,\right]
    \underset{\alpha\to1}{\sim}
    \frac{(2k)!}{2^{k+1}k! (2k-1)} 
    \left| \frac{1}{\alpha-1} \right|^{2k-1}
    \sum_{j=0}^{\lfloor \beta \rfloor } \frac{(-1)^{j}}{j!} (\beta-j)^j e^{\beta-j}.
\end{equation}
\par The asymptotic behavior of the survival probability presented above can be obtained using the direct time-domain approach, as we verify in Appendix~\ref{sec:app_Sinfty}. In contrast, most of the results presented in the remainder of this section are significantly more difficult to obtain using the time-domain approach alone, and here the power of the Laplace-domain framework becomes essential.

\paragraph{Large offset behavior} 
One of the key advantages of the Laplace-domain approach is that it allows us to extract simple closed-form expressions in the limit $\beta\to\infty$. 

\par An important observation concerns the behavior of the survival probability $S_\infty(\beta)$ as $\beta$ increases. The larger the initial offset between the Poisson process and the boundary, the less likely the process is to cross it. We therefore expect, and indeed find, that for $\alpha>1$ the crossing probability $1 - S_\infty(\beta)$ decays exponentially with $\beta$,
\begin{equation}\label{eq:1-Sinfty (beta->inf) res}
    \alpha>1: \quad 
    1 - S_\infty(\beta) \underset{\beta\to\infty}{\asymp}
    \exp\left[
        \beta\left(
            \frac{1}{\alpha}
            + W_{-1}\left[-\frac{1}{\alpha}
            e^{-\frac{1}{\alpha}}\right]
        \right)
    \right],
\end{equation}
where $W_{-1}$ is the secondary real branch of the Lambert function. This exponential suppression implies that for $\alpha>1$ and large $\beta$, almost all trajectories survive, making it unnatural to condition on the rare event of crossing. We therefore focus on the subcritical regime $\alpha<1$, where crossing is the typical outcome.

\par In the subcritical regime ($\alpha<1$), the process eventually crosses the boundary almost surely. We compute the asymptotic expansions of the conditional mean and variance of the first-passage time,
\begin{equation}\label{eq:mean,var~expansion_res}
\alpha<1 :\qquad 
    \mathbb{E}_c\big[\,\tau\,\vert\,\beta\,\big] \underset{\beta\to\infty}{\sim} \frac{\beta}{1-\alpha} + A,
    \qquad
    \mathrm{Var}\big[\,\tau\,\vert\,\beta\,\big] \underset{\beta\to\infty}{\sim}
    \frac{\beta}{(1-\alpha)^3} + B.
\end{equation}
The coefficients $A$ and $B$ are functions depending only on $\alpha$, with explicit expressions provided in \eqref{eq:E_c[tau|beta]_asymp} and \eqref{eq:Var[tau|beta]_asymp}.

\par The linear growth of the mean and the variance with $\beta$ suggests that the probability distribution admits a large deviation form. Specifically, we show that in the limit $\tau\to\infty$ and $\beta\to\infty$ with the ratio held fixed,
\begin{equation}\label{eq:P~LDF_res}
    \alpha<1:\qquad
    \mathbb{P}\big[\,\tau\,\vert\,\beta\,\big] 
    \underset{\beta\to\infty}{\asymp} e^{-\beta \Phi(z)},\qquad
    z = \frac{\alpha \tau}{\beta},
\end{equation}
where the rate function is 
\begin{equation}\label{eq:Phi(z)=res}
    \Phi(z) = \frac{z}{\alpha} - z-1 + (z+1)\log\left[\alpha\left(1+\frac{1}{z}\right)\right].
\end{equation}
The rate function $\Phi(z)$ is non-negative and convex, and vanishes at $z^* = \alpha/(1-\alpha)$, corresponding to the typical crossing time $\tau_\text{typ} = \beta/(1-\alpha)$ from \eqref{eq:mean,var~expansion_res}. Physically, it quantifies the exponential suppression of atypical first-passage times. The asymptotic behavior of $\Phi(z)$ reads:
\begin{equation}\label{eq:Phi(z)=asymptotics res}
    \Phi(z) = 
    \left\{
    \begin{aligned}
        & - \log \frac{z}{\alpha} - 1,
            \quad z\to0,\\
        & \frac{(1-\alpha)^3}{2\alpha^2}\left(z-z^*\right)^2 , 
            \quad z\to z^* = \frac{\alpha}{1-\alpha},\\
        & \left( \log\alpha + \frac{1}{\alpha} - 1\right) z,
            \quad z\to\infty.
    \end{aligned}
    \right.
\end{equation}

\paragraph{Mean first-passage time} 
Finally, we address the general offset $\beta$ and compute the conditional mean first-passage time given in \eqref{eq:Ec[t^k]=def} for $k=1$. Notably, this calculation requires both the time- and Laplace-domain approaches working in tandem. In the subcritical case $\alpha<1$, we obtain
\begin{equation}\label{eq:Ec=subcritical_answer_res}
    \alpha<1:\quad
    \mathbb{E}_{c}\big[\,\tau\,\vert\,\beta\,\big] 
    = 
    \sum_{j=0}^{\lfloor \beta \rfloor}
    \frac{(-1)^{j}}{\alpha^j\, j!}
    \left\{
    \frac{\left( \beta-j\right)^{j} e^{\frac{\beta-j}{\alpha}}}
         {1 + \alpha W_0\left[-\frac{1}{\alpha} e^{-\frac{1}{\alpha}}\right]}
    -\frac{1}{\alpha}
    \int_{0}^{\beta-j}\dd y\, y^{j} e^{\frac{y}{\alpha} }
    \right\},
\end{equation}
while in the supercritical case $\alpha>1$, it reads
\begin{multline}\label{eq:Ec=supercritical_answer_res}
\alpha>1 :\quad 
    \mathbb{E}_c\big[\,\tau\,\vert\,\beta\,\big]
    =\frac{1}{1-S_{\infty}(\beta)}
    \sum_{j=0}^{\lfloor \beta \rfloor}
    \frac{(-1)^{j}}{\alpha^{j+1}\, j!} 
    \Bigg\{   
    \frac{ \left(\beta-j\right)^{j} }{2\, (\alpha-1) } 
        e^{\frac{\beta-j}{\alpha}}
    \\  +
    \left(1-\frac{1}{\alpha}\right)
        \left(\beta-j\right)^{j+1} e^{\frac{\beta-j}{\alpha}} 
    -\int_{0}^{\beta-j} \dd y\, y^{j} e^{\frac{y}{\alpha}}
    \Bigg\},
\end{multline}
with $S_\infty(\beta)$ given in \eqref{eq:Sinfty(beta)=a>1res}.

\par Although \eqref{eq:Ec=subcritical_answer_res} and \eqref{eq:Ec=supercritical_answer_res} are rather cumbersome, they are exact and can be readily evaluated for arbitrary $\alpha$ and $\beta$. Furthermore, in the special case $\beta=0$ they simplify considerably yielding
\begin{align}
    & \alpha<1:\qquad  
   \mathbb{E}_c\big[\,\tau\,\vert\,0\,\big] = \frac{1}{1+\alpha W_0\left[-\frac{1}{\alpha} e^{-\frac{1}{\alpha}}\right]},\\
   & \alpha>1:\qquad \mathbb{E}_c\big[\tau\,\vert\,0\,\big]  = \frac{1}{2} \frac{1}{\alpha-1}.
\end{align}

\par 
We conclude the presentation of the results by noting that the machinery used for the mean extends to higher moments, with the main challenge being technical rather than conceptual. For $\beta=0$, this is particularly straightforward; the second moment, for instance, reads
\begin{equation}
    \alpha<1 : \quad \mathbb{E}_c\left[\,\tau^2\,\vert\,0\,\right] =
    \frac{1}{1+ W_0\left[-\frac{1}{\alpha} e^{-\frac{1}{\alpha}}\right]}
    \frac{2}{\left(1+\alpha W_0\left[-\frac{1}{\alpha} e^{-\frac{1}{\alpha}}\right]\right)^2},
\end{equation}
and
\begin{equation}
    \alpha>1:\qquad \mathbb{E}_{c}\left[\,\tau^2\,\vert\,0\,\right]
        =\frac{1}{6} \frac{2\alpha+1}{(\alpha-1)^3}.
\end{equation}

\section{Two frameworks}\label{sec:TwoFrameworks}
\par 
In this section we present two complementary approaches to studying the first-passage properties of the Poisson process across the linear moving boundary. The first approach operates directly in the time domain and relies on combinatorial identities to obtain an explicit expression \eqref{eq:P=result_time_domain_res} for the probability distribution $\mathbb{P}[\tau\,\vert\,\beta\,]$. The second approach is based on a slight modification of the method introduced in \cite{BM-25}. It maps the original process onto an effective discrete random walk and applies the Pollaczek-Spitzer formula, yielding a representation for the Laplace transform of the first-passage time distribution.

\subsection{Time-domain approach}\label{sec:TimeDomainApproach}
First we use the direct time-domain approach and recover the representations \eqref{eq:S(t,beta)=result_time_domain_res} and \eqref{eq:P=result_time_domain_res} obtained in \cite{P-59}. To analyze the survival probability, it is useful to introduce an intermediate quantity as in \cite{G-66}. Let $P_n(t)$ denote the probability that exactly $n$ jumps have occurred by time $t$ without the process having crossed the boundary:
\begin{equation}\label{eq:Pn(t)=def}
    P_n(t) \equiv \mathrm{Pr}\big[ n \text{ jumps in } (0,t) 
    \text{ and the boundary is not crossed in } (0,t)\big].
\end{equation}
A simple observation shows that if the process has accumulated more than $\lfloor \alpha t + \beta\rfloor$ jumps by time $t$, it must have crossed the boundary at some earlier point, hence
\begin{equation}\label{eq:Tn=def}
    t<T_n: \qquad P_n(t) = 0, \qquad T_n = \max\left\{\frac{n-\beta}{\alpha},0\right\}.
\end{equation}
Physically, $T_n$ represents the time at which the boundary hits the integer value $n$, or equivalently, the earliest time at which the $n$-th jump may occur without causing the crossing. Therefore the survival probability can be written as
\begin{equation}\label{eq:St=sum Pn}
    S(\tau\,\vert\,\beta) = \sum_{n=0}^{\lfloor\alpha \tau+\beta
    \rfloor} P_n(\tau).
\end{equation}
In a similar manner, we now represent the first-passage time distribution $\mathbb{P}[\tau\,\vert\,\beta]$. For the first crossing to occur in the time interval $[\tau,\tau+\delta\tau]$, two conditions must be satisfied: first, there must be no crossings up to time $\tau$, with the process having accumulated exactly $\lfloor \alpha \tau+ \beta \rfloor$ jumps (so that the next jump will definitely cause a crossing); second, the next jump must occur within the interval $\delta \tau$. The probability of a trajectory satisfying the first condition is $P_{\lfloor \alpha \tau+ \beta \rfloor}(\tau)$, while the probability of the next jump occurring within $\delta \tau$ is simply $\delta \tau$, since the Poisson process has unit rate. Therefore, we have
\begin{equation}\label{eq:P=pdvS}
     \mathbb{P}\big[\,\tau\,\vert\,\beta\,\big] =  
     P_{\lfloor \alpha \tau + \beta \rfloor}(\tau).
\end{equation} 
One can easily verify that \eqref{eq:P=pdvS} and \eqref{eq:St=sum Pn} are consistent with \eqref{eq:S(t|beta)=def}. We now turn to computing $P_n(t)$, thereby obtaining the time-domain results for both $\mathbb{P}[\,\tau\,\vert\,\beta\,]$ and $S(\tau\,\vert\,\beta)$.

\begin{figure}[ht]
\includegraphics{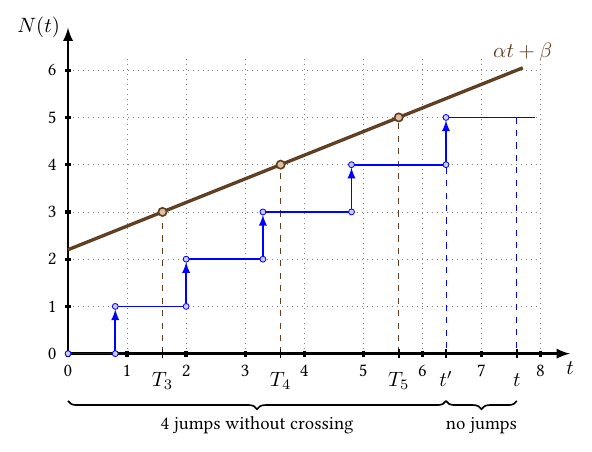}
\caption{Path decomposition corresponding to the recurrence relation \eqref{eq:Pn=recurrence_0} for $n=5$.  
There are 4 jumps without crossing in $(0,t')$, the fifth jump occurs at $t'$, and no jumps occur in the time interval $(t',t)$.
The instances at which the boundary hits an integer value $n$, i.e., the earliest times at which the $n$-th jump may occur without causing a crossing, are given by $T_n$ as in \eqref{eq:Tn=def}.
In this example, the parameters are $\beta=2{.}2$, $\alpha=0.5$ and hence $T_1=T_2=0$, $T_3=1.6$, $T_4=3.6$, and $T_5=5.6$. }\label{fig:recurrence_paths}
\end{figure}

\par The key ingredient is a recurrence relation for $P_n(t)$, which we derive using path decomposition. Let $t'$ denote the time at which the $n$-th jump occurs. For the process to accumulate exactly $n$ jumps by time $t$ without crossing, three statistically independent events must occur:
\begin{enumerate}[noitemsep, nolistsep]
    \item In the time interval $[0,t']$, there are exactly $n-1$ jumps without crossing. The probability of this event is $P_{n-1}(t')$. 
    \item In the infinitesimal interval $[t',t'+\dd t']$, the $n$-th jump occurs. The probability of this event is $\dd t'$.
    \item In the remaining interval $[t'+\dd t', t]$, there are no jumps. The probability of this event is $e^{-(t-t')}$. 
\end{enumerate}
Integrating the product of these three probabilities over all possible jump times $t'$ yields $P_n(t)$. 
The main difference with respect to the analogous recurrence relation for the Poisson process without the boundary \eqref{eq:recurrence_Poisson} is that due to \eqref{eq:Tn=def}, the $n$-th jump can only occur without crossing if $t' > T_n$ (see Fig.~\ref{fig:recurrence_paths}). This constraint leads to the recurrence relation
\begin{equation}\label{eq:Pn=recurrence_0}
    P_n(t) = \int_{T_n}^t \dd t' \, P_{n-1}(t') \; e^{-(t-t')}.
\end{equation}
Factoring out the exponential decay, we obtain an alternative form:
\begin{equation}\label{eq:Cn=recurrence}
    e^{t} P_n(t) = \int_{T_n}^t \dd t'\, e^{t'} P_{n-1}(t');
    \qquad 
    \dv{}{t}\Big[ e^{t}P_n(t) \Big] = e^{t} P_{n-1}(t).
\end{equation}
The initial condition follows from recognizing that $P_0(t)$ is simply the probability of no jumps occurring by time $t$, giving $P_0(t)=e^{-t}$. 

\par 
We emphasize that the boundary enters the recurrence relation \eqref{eq:Pn=recurrence_0} only through the lower limit of integration. Although appearing cosmetic, this small change is actually the source of all technical difficulties. Solving the recurrence relation \eqref{eq:Cn=recurrence} is nontrivial, but it can be shown that
\begin{equation}\label{eq:Pn=closedform}
    P_n(t) = e^{-t} \,
    \frac{(\alpha t + \beta - n)}{\alpha^n} 
        \sum_{j=0}^{\lfloor\beta\rfloor} 
            \frac{(j-\beta)^j(\alpha t + \beta - j)^{n-j-1}}
                 {j!\, (n-j)!}
            \theta(t - T_n),
\end{equation}
where $\theta(t)$ is the Heaviside step function enforcing the constraint \eqref{eq:Tn=def}. Once found, it is straightforward to verify that \eqref{eq:Pn=closedform} satisfies \eqref{eq:Cn=recurrence}.

\par 
As a consistency check, consider the case $n \leq \lfloor\beta\rfloor$. Due to the factor $(n-j)!$, the sum in \eqref{eq:Pn=closedform} is truncated at $n$. In this case the offset is large enough so that $n$ jumps cannot possibly cause a crossing, regardless of when they occur. We therefore expect $P_n(t)$ to simply equal the probability of having $n$ jumps before time $t$, which is given by \eqref{eq:poisson_dist_unit}. To verify this, we use a particular form of Abel's generalization of the binomial theorem \cite{Riordan1968}:
\begin{equation}\label{eq:abel_identity}
    \frac{(x+y+n)^n}{y} =  \sum_{j=0}^n \binom{n}{j} 
    (x+j)^{j} (y + n - j)^{n-j-1}.
\end{equation}
Setting $x = - \beta$ and $y = \alpha t + \beta - n$ in 
\eqref{eq:abel_identity}, we obtain
\begin{equation}\label{eq:sum_simplification}
    \frac{(\alpha t)^{n}}{\alpha t + \beta - n}
    =
    \sum_{j=0}^n 
    \frac{n!}{j!(n-j)!}
    (j-\beta)^j(\alpha t + \beta - j)^{n-j-1}
    .
\end{equation}
Substituting \eqref{eq:sum_simplification} into 
\eqref{eq:Pn=closedform} then yields
\begin{equation}\label{eq:Pn_small_n}
    n \leq \lfloor \beta \rfloor: \quad 
    P_n(t) = \frac{t^n}{n!} e^{-t},
\end{equation}
which matches the Poisson distribution \eqref{eq:poisson_dist_unit}, as expected.

\begin{figure}[ht]
\centering
\includegraphics{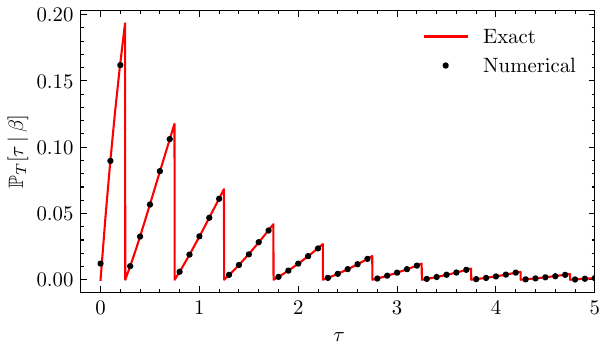}
\caption{First-passage time density $\mathbb{P}\left[\, \tau\,\vert\, \beta\,\right]$ as a function of time for $\beta = 1.5$ and $\alpha = 2$. The red solid line shows the analytical result \eqref{eq:P=result_time_domain}, while black circles correspond to direct numerical simulations obtained by generating $10^{7}$ independent trajectories of the Poisson process. The density exhibits characteristic discontinuities at $T_n = (n-\beta)/\alpha$ for $n > \lfloor\beta\rfloor$.}
\label{fig:fpt_density}
\end{figure}

\par Having obtained the closed form \eqref{eq:Pn=closedform}, we substitute it into \eqref{eq:St=sum Pn} to find 
\begin{equation}\label{eq:S(t,beta)=result_time_domain}
    S(\tau\,\vert\,\beta) = \sum_{n=0}^{\lfloor\alpha \tau+\beta
    \rfloor} 
        \frac{(\alpha \tau + \beta - n)}{\alpha^n} 
        \sum_{j=0}^{\lfloor\beta\rfloor} 
        \frac{(j-\beta)^j(\alpha \tau + \beta - j)^{n-j-1}}
             {j!(n-j)!}e^{-\tau}.
\end{equation}
This is exactly the result obtained by Pyke in \cite{P-59}. While Pyke's derivation was formulated in a different way, it relies on essentially the same combinatorial identities as the presented approach. 
\par Similarly, substituting \eqref{eq:Pn=closedform} into \eqref{eq:P=pdvS} we immediately obtain the closed form expression for the probability density of the first-passage time
\begin{equation}\label{eq:P=result_time_domain}
\mathbb{P}\left[\,\tau\,\vert\,\beta\,\right] 
    = \frac{(\alpha \tau + \beta - \lfloor\alpha \tau + \beta
            \rfloor)}
           {\alpha^{\lfloor\alpha \tau+\beta\rfloor}} 
    \sum_{j=0}^{\lfloor\beta\rfloor} 
        \frac{(j-\beta)^j(\alpha \tau + \beta - j)^{\lfloor\alpha 
              \tau+\beta\rfloor-j-1}}
             {j!(\lfloor\alpha \tau+\beta\rfloor - j)!}e^{-\tau}.
\end{equation}
We emphasize that \eqref{eq:S(t,beta)=result_time_domain} and \eqref{eq:P=result_time_domain} are valid for arbitrary offsets $\beta\ge0$ and drift velocities $\alpha>0$.  An example of the first-passage time density is shown in Fig.~\ref{fig:fpt_density}, where the analytical prediction \eqref{eq:P=result_time_domain} is compared against direct numerical simulations of the Poisson process. 

\par
\begin{figure}[h]
    \includegraphics{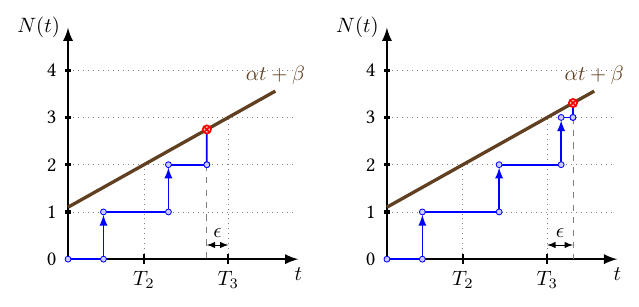}
    \caption{An example of a configuration with the crossing at  $\tau=T_3-\epsilon$ (left) and at $\tau=T_3+\epsilon$ (right). A crossing at $\tau=T_3+\epsilon$ would require two jumps within the interval $(T_3,T_3+\epsilon)$, which becomes impossible as $\epsilon\to0$, hence the different limits in \eqref{eq:P[tn-epsilon]=} and \eqref{eq:P[tn+epsilon]=}. }
    \label{fig:paths_discontinuity}
\end{figure}

 The explicit appearance of the floor function in \eqref{eq:P=result_time_domain} means that the density has sharp discontinuities at the points $\tau=\frac{n-\beta}{\alpha}$ for $n> \lfloor \beta \rfloor$. These are exactly the instances $T_n$ given in \eqref{eq:Tn=def}, i.e., the earliest times at which the $n$-th jump may occur without causing the crossing. The discontinuities can be characterized by 
\begin{align}
\label{eq:P[tn-epsilon]=}
    \lim_{\epsilon\to 0^+}\mathbb{P}[\,T_n - \epsilon \,\vert\, 
    \beta\,] 
    &= \frac{1}{\alpha^{n-1}}\sum_{j=0}^{\lfloor\beta\rfloor} 
       \frac{(j-\beta)^j(n-j)^{n-j-1}}{j!(n-j)!}
       e^{-(n-\beta)/\alpha}, \\
    \label{eq:P[tn+epsilon]=}
    \lim_{\epsilon\to 0^+}
    \mathbb{P}[\,T_n + \epsilon\,\vert\,\beta\,] &= 0.
\end{align}
While initially surprising, this discontinuous behavior has a simple physical explanation: there cannot be a first-passage event immediately after $T_n$, i.e., the instant at which the boundary hits the integer $n$ (see Fig.~\ref{fig:recurrence_paths}). Indeed, a first passage at $\tau = T_n + \epsilon$ would require the $n$-th and $(n+1)$-th jumps to both occur within the interval $(T_n, T_n+\epsilon)$, which becomes impossible as $\epsilon \to 0$ (see Fig.~\ref{fig:paths_discontinuity}). This explains the result in \eqref{eq:P[tn+epsilon]=}. 

\par Although the representation \eqref{eq:P=result_time_domain} is exact, it is not always convenient for extracting physical quantities of interest. As an illustration, let us examine the first-passage probability density at $\tau=0$. From a physical perspective, if $\beta<1$, a single jump of the Poisson process suffices to cross the boundary. Since the process has unit rate, the probability density for a jump in an infinitesimal time interval is unity. Conversely, if $\beta\ge 1$, the boundary crossing requires at least two jumps, and the density at $\tau=0$ must vanish. The first-passage probability density therefore must satisfy:
\begin{equation}\label{eq:P[0]=answer}
    \lim_{\tau\to0}\mathbb{P}[\,\tau\,\vert\,\beta\,] = 
    \begin{cases}
        1, & \beta<1,\\
        0, & \beta\ge 1.
    \end{cases}
\end{equation}
At the same time, this behavior is not immediately apparent if we simply substitute $\tau=0$ in \eqref{eq:P=result_time_domain}, and computing it requires a slightly more careful approach. For example, we can first change the upper limit in the sum from $\lfloor \beta \rfloor$ to $\lfloor \alpha\tau+\beta\rfloor$, which does not change the result as $\tau \to 0$. We can then compute the sum using the identity \eqref{eq:abel_identity} with $n=\lfloor \alpha\tau+\beta\rfloor$, $x=-\beta$, and $y=\alpha\tau+\beta-\lfloor \alpha\tau+\beta\rfloor$:
\begin{equation}\label{eq:abel_identity_t=0}
\sum_{j=0}^{\lfloor\alpha\tau + \beta\rfloor} 
\frac{(j-\beta)^j(\alpha \tau + \beta - j)^{\lfloor\alpha 
      \tau+\beta\rfloor-j-1}}
     {j!(\lfloor\alpha \tau+\beta\rfloor - j)!}
= \frac{1}{\lfloor \alpha\tau+\beta\rfloor!} 
\frac{(\alpha \tau)^{\lfloor \alpha\tau + \beta\rfloor}}
     {\alpha\tau+\beta - \lfloor \alpha\tau+\beta\rfloor}.
\end{equation}
The limiting behavior is then given by
\begin{equation}\label{eq:P[0]=answer_exact}
    \lim_{\tau\to0}\mathbb{P}[\,\tau\,\vert\,\beta\,] = 
    \lim_{\tau\to0} 
        \frac{\tau^{\lfloor\alpha\tau+\beta\rfloor}}
             {\lfloor \alpha\tau+\beta\rfloor!}e^{-\tau}
    = 
    \begin{cases}
        1, & \beta<1,\\
        0, & \beta\ge 1,
    \end{cases}
\end{equation}
which agrees with \eqref{eq:P[0]=answer}. 

\par 
We presented this calculation both as a consistency check and as an illustration that even though \eqref{eq:P=result_time_domain} is explicit, using it for computations is not always straightforward. For such a simple question as the density at $\tau=0$ we were able to perform the calculation without much effort, but the technical difficulties magnify significantly when considering conditional moments \eqref{eq:Ec[t^k]=def} or deriving the asymptotic behavior of $S(\tau\,\vert\,\beta)$, as will be illustrated later.

\subsection{Laplace-domain approach}\label{sec:LaplaceDomainApproach}
We now turn to an alternative approach based on the technique from \cite{BM-25}. The aim is to recover the representation \eqref{eq:hat(S)=explicit_res} of the double Laplace transform $\hat{S}(\rho\,\vert\,\lambda)$. In contrast to the direct time-domain approach, here it is more convenient to work with the first-passage probability density $\mathbb{P}\left[\,\tau\,\vert\,\beta\,\right]$ rather than the survival probability $S\left(\tau\,\vert\,\beta\right)$. These two quantities are related via \eqref{eq:S(t|beta)=def} and \eqref{eq:hatS=1-hatQ}.

\par Consider the distance $X(t)$ between the Poisson process and the moving boundary. When no jumps occur, this distance increases linearly with velocity $\alpha$ as the boundary advances. Whenever a jump occurs, the distance suddenly decreases by one unit. Therefore, $X(t)$ evolves according to
\begin{equation}\label{eq:X(t)=dynamics_Poisson}
    X(t) = \beta + \alpha t - N(t),
\end{equation}
where $N(t)$ is the number of jumps that have occurred up to time $t$, and $\beta$ is the initial distance. The first-passage event of the original problem (the Poisson process crossing the boundary) thus corresponds to $X(t)$ becoming negative.

\par 
The process \eqref{eq:X(t)=dynamics_Poisson} can be viewed as the dual counterpart to the one studied in \cite{BM-25}. In the setup of this paper, that problem would correspond to first-passage with respect to a lower boundary (rather than the upper boundary considered here). To make this parallel more apparent, let us briefly consider a more general version of~\eqref{eq:X(t)=dynamics_Poisson}.

\paragraph{General formalism} 
Suppose, for a moment, that the jumps occur not according to a Poisson process but rather according to a renewal process $n(t)$, i.e., the inter-jump time intervals $t_j$ are independent random variables drawn from a distribution $p(t)$. Furthermore, suppose that the jump amplitudes $M_j$ are not fixed to one, but are instead independent random variables drawn from a distribution $q(M)$. Then the evolution of the process can be written as
\begin{equation}\label{eq:X(t)=dynamics}
    X(t) = \beta + \alpha t - \sum_{j=1}^{n(t)} M_j.
\end{equation}
Note that the dynamics \eqref{eq:X(t)=dynamics} reduces to \eqref{eq:X(t)=dynamics_Poisson} when $p(t)=e^{-t}$ and $q(M)=\delta(M-1)$.

\par Moreover, instead of focusing solely on the first-passage time distribution $\mathbb{P}\left[\,\tau\,\vert\,\beta\,\right]$, let us consider the joint distribution $\mathbb{P}\left[\,\tau,n\,\vert\,\beta\,\right]$ of the first-passage time $\tau$ and the index $n$ of the jump that causes the crossing. Since $X(t)$ increases monotonically between jumps, the crossing can only occur at a jump time. The marginal distribution is then recovered by summing over $n$:
\begin{equation}\label{eq:P_marginal}
    \mathbb{P}\left[\,\tau\,\vert\,\beta\,\right] = 
    \sum_{n=1}^{\infty} \mathbb{P}\left[\,\tau,n\,\vert\,\beta\,\right].
\end{equation}

\par
The first-passage properties of \eqref{eq:X(t)=dynamics} can be studied using an approach that closely follows that of \cite{BM-25}. The general idea is to first apply a mapping technique introduced in \cite{MDMS-20} (see also \cites{MDMS-20b,MMV-24,SBEM-25}) that transforms the process into an effective random walk. Then, applying the generalization \cite{I-94} of the Pollaczek-Spitzer formula \cites{P-52,S-56,S-57}, we obtain a closed-form representation for the Laplace transform of the joint distribution $\mathbb{P}\left[\,\tau,n\,\vert\,\beta\,\right]$.

\begin{figure}[ht]
\includegraphics{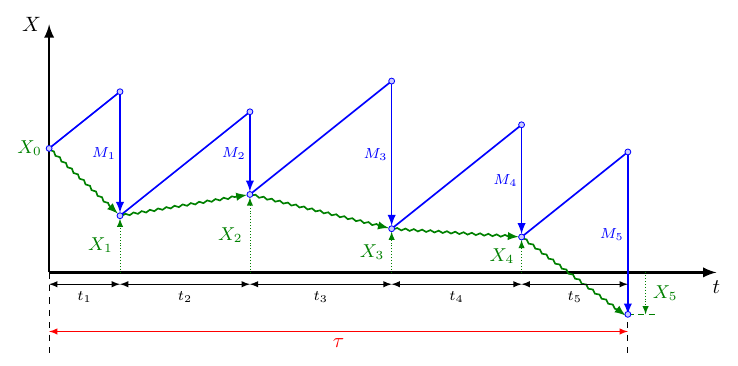}
\caption{An example trajectory of the process \eqref{eq:X(t)=dynamics} (blue) and the corresponding effective random walk (green). 
The first passage happens at time $\tau$ and is caused by the fifth jump ($n=5$). }\label{fig:discrete_time_walk}
\end{figure}

\par We begin by constructing an effective random walk. Let $X_j$ denote the coordinate of $X(t)$ immediately after the $j$-th jump (see Fig.~\ref{fig:discrete_time_walk}). Since both the waiting times $t_j$ and the jump amplitudes $M_j$ are random variables, the difference between consecutive positions,
\begin{equation}\label{eq:X_discrete_walk}
    X_{j+1} = X_{j} + \eta_j,\qquad 
    \eta_j = \alpha t_j - M_j,
\end{equation}
defines a discrete-time random walk with random increments  $\eta_j$. Starting at $X_0=\beta$, this walk envelops the  trajectory of \eqref{eq:X(t)=dynamics} from below. Since $X(t)$  increases monotonically between jumps, the continuous process can only cross the origin at a jump time. Therefore, the first-passage event of the continuous process corresponds to the first index $k$ such that $X_k < 0$.

\par 
To complete the mapping, we must specify the probability distribution of the increments $\eta_j$. While \eqref{eq:X_discrete_walk} defines $\eta_j$ as a difference of two random variables, naively using the distributions $p(t)$ and $q(M)$ to characterize $\eta_j$ discards crucial temporal information. Such an approach correctly tracks whether a crossing occurs and determines the number of jumps $n$ required, but it loses all information about \emph{when} the crossing occurs, failing to capture the first-passage time $\tau$. A more careful treatment is therefore needed.

\par 
The joint probability distribution $\mathbb{P}\left[\,\tau,n\,\vert\,\beta\,\right]$ is obtained by integrating over all possible sequences of waiting times $t_j$ and jump amplitudes $M_j$ that lead to a crossing at the $n$-th jump. To ensure the crossing occurs precisely at this jump, we impose the constraint that the first $(n-1)$ positions remain positive, $X_j > 0$ for $j=1,\ldots,n-1$, while the $n$-th position becomes negative, $X_n < 0$. This constraint is enforced by step functions $\theta(X_j)$ for $j < n$ and $\theta(-X_n)$. The joint distribution can then be written as
\begin{multline}\label{eq:P=intintint_explicit}
    \mathbb{P}\left[\,\tau,n\,\vert\,\beta\,\right] = 
    \int_{0}^{\infty} \dd M_1 \ldots \dd M_n
    \int_{0}^{\infty} \dd t_1\ldots\dd t_n
    \\
    \times
    \delta\left(\tau - \sum_{j=1}^{n} t_j\right)
    \theta(-X_{n}) 
    \prod_{j=1}^{n-1} \theta(X_j) 
    \prod_{j=1}^{n} p(t_j) q(M_j),
\end{multline}
where the $\delta$-function fixes the first passage time to be $\tau$, and $X_j$ according to \eqref{eq:X_discrete_walk} are given by
\begin{equation}\label{eq:X_j=def_sum}
    X_j = \beta + \sum_{k=1}^{j}(\alpha t_k - M_k)
        = \beta + \sum_{k=1}^{j} \eta_k.
\end{equation}
Performing a Laplace transform with respect to $\tau$ in \eqref{eq:P=intintint_explicit} we arrive at
\begin{multline}\label{eq:P=intintint_explicit_LT1}
    \int_{0}^{\infty} \dd \tau\, e^{-\rho\tau} \, \mathbb{P}\left[\,\tau,n\,\vert\,\beta\,\right] = 
    \int_{0}^{\infty} \dd M_1 \ldots \dd M_n
    \int_{0}^{\infty} \dd t_1\ldots\dd t_n
    \\
    \times
    \theta(-X_{n}) 
    \prod_{j=1}^{n-1} \theta(X_j) 
    \prod_{j=1}^{n} e^{-\rho t_j}p(t_j) q(M_j).
\end{multline}
Then we change the variable of integration $M\mapsto\eta$ with $\eta_j = \alpha t_j - M_j$, so that  $X_j$ is now a function of $\{\eta_k\}_{k=1,\ldots,j}$ as given in \eqref{eq:X_j=def_sum}, 
\begin{multline}\label{eq:P=intintint_explicit_LT1_eta}
    \int_{0}^{\infty} \dd \tau\, e^{-\rho\tau} \, \mathbb{P}\left[\,\tau,n\,\vert\,\beta\,\right] = 
    \int_{-\infty}^{\infty} \dd \eta_1 \ldots \dd \eta_n\;
    \theta(-X_{n}) 
    \prod_{j=1}^{n-1} \theta(X_j) 
    \\
    \times
    \int_{0}^{\infty} \dd t_1\ldots\dd t_n
    \prod_{j=1}^{n} e^{-\rho t_j}p(t_j) q(\alpha t_j - \eta_j).
\end{multline}
The representation \eqref{eq:P=intintint_explicit_LT1_eta} is not particularly revealing as it stands, but one can recognize it as the survival probability of the effective discrete-time random walk using the trick introduced in~\cite{MDMS-20}. To do this, we define the notation
\begin{equation}\label{eq:f(eta;rho)=def}
    f(\eta;\rho) = \frac{1}{c(\rho)} \int_{0}^{\infty} \dd t\, e^{-\rho t} p(t) q(\alpha t - \eta),\qquad
    c(\rho) = \int_{0}^{\infty}\dd t\, e^{-\rho t} p(t).
\end{equation}
An important observation is that $f(\eta;\rho)$ is normalized to one,
\begin{equation}
    \int_{-\infty}^{\infty} \dd \eta\, f(\eta;\rho)
    = \frac{1}{c(\rho)} \int_{-\infty}^{\infty} \dd \eta\, 
    \int_{0}^{\infty} \dd t\, e^{-\rho t} p(t) q(\alpha t - \eta).
\end{equation}
To verify this, we change variables $\eta\to M$, where $M=\alpha t - \eta$, and use the fact that $q(M)$ vanishes for $M<0$:
\begin{equation}
    \int_{-\infty}^{\infty} \dd \eta\, f(\eta;\rho)
    = \frac{1}{c(\rho)} \int_{0}^{\infty} \dd t\, e^{-\rho t} p(t) 
        \int_{0}^{\infty} \dd M\, q(M)  = 1.
\end{equation}
We can then rewrite \eqref{eq:P=intintint_explicit_LT1_eta} as
\begin{multline}\label{eq:P=intintint_explicit_LT2}
    \int_{0}^{\infty} \dd \tau\, e^{-\rho\tau} \, \mathbb{P}\left[\,\tau,n\,\vert\,\beta\,\right] 
    =
    \big[c(\rho)\big]^{n} 
    \int_{-\infty}^{\infty}\dd \eta_1\ldots\dd\eta_n
    \\\times
    \theta(-X_{n}) 
    \prod_{j=1}^{n-1} \theta(X_j) \;
    \prod_{j=1}^{n} f(\eta_j;\rho).
\end{multline}
Since $f(\eta;\rho)$ is positive and normalized, it can be interpreted as a probability density. The integral in \eqref{eq:P=intintint_explicit_LT2} is then nothing but the probability that the discrete-time random walk $X_j$ with increments $\eta$ drawn from $f(\eta;\rho)$ crosses the origin exactly at the $n$-th step. This probability can be computed using a generalization of the Pollaczek-Spitzer formula as explained in \cite{BM-25}. 

\par 
Before presenting the result, let us make a brief comment on the role of the parameter~$\rho$. When $\rho=0$, we have $c(0)=1$ and
\begin{equation}
    f(\eta;0) = \int_{0}^{\infty} \dd t\, p(t) q(\alpha t - \eta)
    = \int_{0}^{\infty} \dd M \int_{0}^{\infty} \dd t\, 
     p(t) q(M) \delta(\eta - \alpha t + M),
\end{equation}
which is the probability distribution of $\eta = \alpha t - M$ as the difference of two random variables with distributions $p(t)$ and $q(M)$. In this case, temporal information is lost: each step consists of a waiting period (which increases the distance by $\alpha t$) followed by a jump (which decreases it by $M$), and many different combinations of waiting time and jump size can produce the same increment $\eta$. The parameter $\rho$ modifies the relative weight of longer vs. shorter waiting times, allowing us to distinguish between these combinations and thereby preserve the temporal information.

\par Once the connection \eqref{eq:P=intintint_explicit_LT2} between the first-passage probability distribution of the process \eqref{eq:X(t)=dynamics} and the survival probability of the effective random walk is established, we apply the generalization of the Pollaczek-Spitzer formula to find the closed-form representation of the Laplace transform of $\mathbb{P}\left[\,\tau,n\,\vert\,\beta\,\right]$:
\begin{equation}\label{eq:hat(Q)(rho,s|lambda)=def}
    \hat{\mathcal{Q}}\left(\rho,s\,\vert\,\lambda\right) \equiv 
    \int_{0}^{\infty} \dd \tau\, e^{-\rho \tau}
    \int_{0}^{\infty} \dd \beta\, e^{-\lambda \beta}
    \sum_{n=1}^{\infty} s^{n}\,
    \mathbb{P}\left[\,\tau,n\,\vert\,\beta\,\right].       
\end{equation}  
From \eqref{eq:hat(Q)=def} and \eqref{eq:P_marginal}, it is clear that
\begin{equation}\label{eq:Q(rho,s|lambda)=Q(rho|lambda)}
    \hat{\mathcal{Q}}(\rho,s\,\vert\,\lambda)\Big|_{s=1} = \hat{Q}(\rho\,\vert\,\lambda). 
\end{equation}
The derivation of the explicit representation for $\hat{\mathcal{Q}}(\rho,s\,\vert\,\lambda)$  follows along the lines of \cite{BM-25}; we therefore present only the result here. Specifically, we find that the triple Laplace transform is given by
\begin{equation}\label{eq:hatQ=PollSpitzer}
    \hat{\mathcal{Q}}(\rho,s\,\vert\,\lambda) = 
    \frac{1}{\lambda} - 
    \frac{1 - s\,c(\rho)}{\lambda}
     \phi^{-}(\lambda;\rho,s) \,
    \phi^{+}(0;\rho,s).
\end{equation}
The functions $\phi^\pm(\lambda;\rho,s)$ are defined as
\begin{equation}\label{eq:phi^pm=def}
    \phi^{\pm}(\lambda;\rho,s) = 
    \exp\left[
        -\frac{1}{2\pi} \int_{-\infty}^{\infty}\dd k\, \frac{1}{\lambda\pm\ii k} \log\left[1 - s c(\rho) F(k;\rho)\right]
    \right],
\end{equation}
where $F(k;\rho)$ is the Fourier transform of $f(\eta;\rho)$ given by \eqref{eq:f(eta;rho)=def}. In terms of the inter-jump times probability density $p(t)$ and jump amplitudes density $q(M)$, it reads
\begin{equation}\label{eq:F(k;rho)=p,q}
    F(k;\rho) = \frac{1}{c(\rho)}
     \int_{0}^{\infty} \dd t\, e^{-\rho t + \ii k \alpha t} p(t)
    \int_{0}^{\infty} \dd  M\, e^{-\ii k M} q(M).
\end{equation}
We emphasize that the representation \eqref{eq:hatQ=PollSpitzer} holds for arbitrary distributions $p(t)$ and $q(M)$. Despite appearing intimidating, it permits the extraction of various asymptotic properties via analytic continuation techniques similar to those in \cite{B-25}. We have presented this general result here primarily to demonstrate that the approach can be applied to a much broader class of systems than the specific process considered in the present paper (see \cite{ACEK-14,BCP-25} for the case in which jumps and waiting times are correlated). However, we will not pursue these generalizations any further. Instead, we conclude our detour into the general case and return to the problem at hand.

\paragraph{Poisson process case}
Now we use the representation \eqref{eq:hatQ=PollSpitzer} to compute the Laplace transform of the first-passage time distribution for the Poisson process with respect to a linear moving boundary. This is done by opting for the inter-jump times to follow the exponential distribution $p(t)=e^{-t}$ so that the jumps occur according to a Poisson process and for the jump amplitudes to be fixed to unity, i.e., $q(M)=\delta(M-1)$. With this choice, the effective random walk characterized by \eqref{eq:f(eta;rho)=def} and \eqref{eq:F(k;rho)=p,q} reduces to
\begin{equation}\label{eq:F(k;rho)=Poisson}
    F(k;\rho) = \frac{\rho+1}{\rho+1-\ii\alpha k} \, e^{-\ii k},
    \qquad
    c(\rho) = \frac{1}{\rho+1}.
\end{equation}
In this case, the integrals in \eqref{eq:phi^pm=def} can be computed via standard residue calculus techniques (we refer the reader to Appendix~\ref{sec:app_Q_Laplace} for details). The resulting expression reads:
\begin{equation}\label{eq:hat(Q)(rho,s|lambda)=explicit}
    \hat{\mathcal{Q}}(\rho,s\,\vert\,\lambda) =
    \frac{1}{\lambda} 
    - \frac{1}{\lambda}\frac{\rho+1-s}{\rho+1-\alpha\lambda - s \, e^{-\lambda}}
    \left(1 - \frac{\alpha\lambda}{\rho+1 + \alpha W_0\left[-s \frac{1}{\alpha}e^{-\frac{\rho+1}{\alpha}}\right]}\right),
\end{equation}
where $W_0$ denotes the principal branch of the Lambert W-function. Recalling \eqref{eq:Q(rho,s|lambda)=Q(rho|lambda)}, we immediately find that
\begin{equation}\label{eq:hat(Q)=explicit}
    \hat{Q}(\rho\,\vert\,\lambda) = 
    \frac{1}{\lambda} -
    \frac{1}{\lambda}
    \frac{\rho}{\rho+1-\alpha\lambda - e^{-\lambda}}
    \left(
        1 - \frac{\alpha \lambda}
                 {\rho+1 + \alpha W_0\left[
                     -\frac{1}{\alpha} e^{-\frac{\rho+1}{\alpha}}
                 \right] }
    \right).
\end{equation}
Once \eqref{eq:hat(Q)=explicit} is known, one can readily obtain the double Laplace transform of the survival probability using \eqref{eq:hatS=1-hatQ}, arriving at
\begin{equation}\label{eq:hat(S)=explicit}
    \hat{S}(\rho\,\vert\,\lambda)= 
    \frac{1}{\lambda}
    \frac{1}{\rho + 1 - \alpha\lambda - e^{-\lambda} }
    \left(1 - \frac{\lambda}{\frac{\rho+1}{\alpha} + W_0\left[ 
    -\frac{1}{\alpha} e^{-{\frac{\rho+1}{\alpha}}} 
    \right]} \right).
\end{equation}
This is exactly the result stated in \eqref{eq:hat(S)=explicit_res}. 

\par We emphasize that although the resulting expression is the same as the one presented in \cite{BD-57} and both derivations rely on the Pollaczek-Spitzer formula, the path we took to find \eqref{eq:hat(S)=explicit} is different, and we believe it to be more transparent and versatile, since the connection to the theory of discrete-time random walks is explicit.

\paragraph{<<Tweedledum and Tweedledee>>}
Let us now pause for a moment to reflect on the two approaches developed above. Both methods tackle the same problem but employ fundamentally different machinery: the time-domain approach leverages combinatorial identities and recurrence relations, while the Laplace-domain approach relies on complex analysis.

\par Having derived the survival probability through both approaches~--- the time-domain method yielding \eqref{eq:S(t,beta)=result_time_domain} and the Laplace-domain method yielding \eqref{eq:hat(S)=explicit}~--- these representations must be equivalent. Writing this equivalence explicitly reveals a highly nontrivial integral identity:
\begin{multline}\label{eq:equivalence_of_two_approaches} 
    \int_{0}^\infty\! \dd\beta
    \int_{0}^\infty\! \dd \tau\, e^{-\lambda\beta-(\rho+1)\tau} 
    \sum_{n=0}^{\lfloor\alpha \tau+\beta\rfloor} 
            \frac{(\alpha \tau + \beta - n)}{\alpha^n} 
    \sum_{j=0}^{\lfloor\beta\rfloor} 
            \frac{(j-\beta)^j(\alpha \tau + \beta - j)^{n-j-1}}{j!(n-j)!}
    \\
    =\frac{1}{\lambda\left( 1 + \rho - \alpha\lambda - e^{-\lambda} \right)}
    \left(1 - \frac{\alpha \lambda}{1+\rho + \alpha W_0\left[ -\frac{1}{\alpha} e^{-{\frac{\rho+1}{\alpha}}} \right]} \right) 
    .
\end{multline}
An analogous identity follows by comparing the time- and Laplace-domain representations of the first-passage probability density, as given in \eqref{eq:P=result_time_domain} and \eqref{eq:hat(Q)=explicit}.

\par The identity \eqref{eq:equivalence_of_two_approaches} is far from obvious, and discovering it without guidance from either approach would require an extraordinary stroke of luck. Moreover, even once found, verifying it is a challenging task. We perform this verification for the limit $\lambda\to\infty$, which corresponds to the zero offset $\beta=0$, in Section~\ref{subsec:equivalence}. A direct proof for the general $\lambda$ remains to be found.

\par The two approaches complement each other: the time-domain technique yields an explicit closed form for the probability distribution that can be readily computed numerically for any given $\tau$ and $\beta$, but is ill-suited for computing moments or performing asymptotic analysis. The Laplace-domain approach, by contrast, is naturally suited for these tasks. Most importantly, having both results at hand allows us to obtain expressions that would be difficult or impossible to derive from either method alone.

\par Although most of the computations presented in the remainder of the paper rely on Laplace-domain techniques, this should not mislead the reader into thinking that the Laplace-domain approach is somehow superior. We implicitly leverage intuition from the time-domain approach in many situations. The main examples are the exact representations for the mean first-passage time and the survival probability at infinite time for general offset $\beta$, which we derive in Section~\ref{sec:general_offset}.

\section{Extreme cases}\label{sec:Extreme cases}
\par Before analyzing the general case, we focus on two instructive special scenarios: the zero offset limit ($\beta=0$) and the infinite offset limit ($\beta\to\infty$). These extreme cases serve to validate the two approaches developed in the previous section. Moreover, they provide clear insights into essential features of the problem without the technical complications that arise in the general case. 

\par Before proceeding with the computations, let us introduce notation for the Laplace transform of $\mathbb{P}\left[\,\tau\,\vert\,\beta\,\right]$ with respect to $\tau$ only:
\begin{equation}\label{eq:Q(rho|beta)=def}
    Q(\rho\,\vert\,\beta) \equiv
    \int_0^{\infty} \dd \tau\, e^{-\rho\tau} \,
    \mathbb{P}\left[\,\tau\,\vert\,\beta\,\right].
\end{equation}
This differs from the double Laplace transform $\hat{Q}(\rho\,\vert\,\lambda)$ defined in \eqref{eq:hat(Q)=def}.

\subsection{Zero offset}
\par We start with the zero offset case in which $\beta=0$. In this scenario the expression for the survival probability \eqref{eq:S(t,beta)=result_time_domain} that was obtained using the time-domain approach simplifies significantly to
\begin{equation}\label{eq:S(t|0)=result_time_domain}
    S(\tau\,\vert\,0) = \sum_{n=0}^{\lfloor\alpha \tau\rfloor} 
        \left(\tau - \frac{n}{\alpha}\right) 
        \frac{\tau^{n-1}}{n!} e^{-\tau}.
\end{equation} 
Similarly, the first-passage probability density \eqref{eq:P=result_time_domain} reduces to
\begin{equation}\label{eq:P[tau|0]=result_time_domain}
\mathbb{P}[\,\tau\,\vert\,0\,] 
    =
    \frac{1}{\alpha \tau}   
    \frac{\tau^{\lfloor\alpha \tau\rfloor}}{\lfloor\alpha \tau\rfloor!}
    (\alpha \tau - \lfloor\alpha \tau\rfloor) e^{-\tau}.
\end{equation}
These expressions are explicit and well-suited for numerical evaluation. However, extracting analytic properties, such as moments or asymptotic behavior, from them is not particularly straightforward~--- the floor function makes direct integration impractical. This is where the Laplace-domain approach enters the stage.

\par To extract $\beta=0$ behavior of $\mathbb{P}\left[\,\tau\, \vert\,0\,\right]$ from the Laplace transform \eqref{eq:hat(Q)=explicit} we need to study the $\lambda\to\infty$ behavior of $\hat{Q}(\rho\,\vert\,\lambda)$. A simple calculation yields
\begin{equation}\label{eq:hatQ=lambda->infty}
    \hat{Q}(\rho\,\vert\,\lambda) 
    \underset{\lambda\to\infty}{=} \frac{1}{\lambda}
    \left(
        1 - \frac{\rho}{1 + \rho + \alpha \, W_0\left[- \frac{1}{\alpha} e^{ - \frac{\rho+1}{\alpha} }\right]}
    \right) 
    + O\left(\frac{1}{\lambda^2}\right),
\end{equation}
from which we deduce that the single Laplace transform \eqref{eq:Q(rho|beta)=def} is given by
\begin{equation}\label{eq:hatQ(rho,0)=}
    Q(\rho\,\vert\,0) = 
    \int_0^\infty \dd \tau \, e^{-\rho \tau}\,
    \mathbb{P}\left[\, \tau\,\vert\,0\,\right] 
    = 1 -  \frac{\rho}
               {1 + \rho + \alpha \, W_0\left[- \frac{1}{\alpha} e^{ - \frac{\rho+1}{\alpha} }\right]  } .
\end{equation}
Similarly, for the Laplace transform of the survival probability we find
\begin{equation}\label{eq:hatS(rho,0)=}
    \int_0^\infty \dd \tau\,  e^{-\rho \tau} S(\tau\,\vert\,0) 
    = \frac{1}
           {1 + \rho + \alpha \, W_0\left[- \frac{1}{\alpha} e^{ - \frac{\rho+1}{\alpha} }\right]  } .
\end{equation}
Recall that $W_0(z)$ denotes the principal branch of the Lambert function.

\par At first glance, the Laplace-domain expressions \eqref{eq:hatQ(rho,0)=} and \eqref{eq:hatS(rho,0)=} appear no simpler than their time-domain counterparts, as the Lambert function $W_0$ is itself transcendental. However, as we shall see, this representation has a crucial advantage: it makes moments and asymptotic behavior readily accessible.

\par In what follows, we address two questions. First, we establish the equivalence between the seemingly disparate representations \eqref{eq:S(t|0)=result_time_domain} and \eqref{eq:hatS(rho,0)=}. This serves as a nontrivial check that both approaches yield consistent results and also provides the intuition that we will use later in Section~\ref{sec:general_offset}. 
Second, we use the Laplace form \eqref{eq:hatQ(rho,0)=} to derive closed form expressions for conditional moments \eqref{eq:Ec[t^k]=def}~--- a calculation that would be exceedingly difficult starting from \eqref{eq:P[tau|0]=result_time_domain} alone.

\subsubsection{Equivalence of the approaches}\label{subsec:equivalence}
\par We first prove that \eqref{eq:hatS(rho,0)=} is indeed the Laplace transform of \eqref{eq:S(t|0)=result_time_domain}. Remarkably, the inverse problem, i.e.,  showing that 
\eqref{eq:S(t|0)=result_time_domain} is the inverse Laplace transform of \eqref{eq:hatS(rho,0)=}, turns out to be more tractable.

\par While in principle the inversion can be performed via residue calculus by directly analyzing the singularity structure of \eqref{eq:hatS(rho,0)=} in the $\rho$-plane, the result does not immediately reduce to \eqref{eq:S(t|0)=result_time_domain}. We therefore employ an alternative strategy based on series expansions of the Lambert function. In particular, we rely on the following expansion (see Appendix~\ref{sec:app_LambertW}):
\begin{equation}\label{eq:W0^j=expansion}
    \Big[ W_0(x) \Big]^j 
    = -
    j \sum_{n=j}^{\infty}  \frac{(-n)^{n-j-1}}{(n-j)!}  x^n,
    \quad j\ge 1,
\end{equation}
which is valid around the origin with radius of convergence equal to $1/e$.

\par 
Note that due to \eqref{eq:W0^j=expansion}, the Lambert function in \eqref{eq:hatS(rho,0)=} decays exponentially fast as $\rho\to\infty$. Treating it as a small parameter, we obtain the asymptotic series representation
\begin{equation}\label{eq:LTS(0)=sum_auxiliary}
\int_0^\infty \dd \tau\,  e^{-\rho \tau} S(\tau\,\vert\,0)
    = 
    \frac{1}{1+\rho}
    \sum_{j=0}^\infty
    \frac{(-\alpha)^j}{(1+\rho)^j} 
    \left( W_0\left[- \frac{1}{\alpha} e^{ - \frac{\rho+1}{\alpha} }\right] \right)^{j}.
\end{equation}
Then, applying \eqref{eq:W0^j=expansion} yields
\begin{equation}\label{eq:LTS(0)=double_sum}
    \int_{0}^{\infty} \dd \tau\,
    e^{-\rho\tau}
    S(\tau\,\vert\,0)
    = 
    \frac{1}{1+\rho}
    +
    \sum_{j=1}^\infty
    \frac{j \, \alpha^{j-n}}{(1+\rho)^{j+1}} 
    \sum_{n=j}^{\infty} \frac{n^{n-j-1}}{(n-j)!}  
    e^{-\frac{\rho+1}{\alpha} n} .
\end{equation}
These infinite series converge, which makes the rearrangement of terms possible. By keeping track of the powers of~$e^{-\frac{\rho+1}{\alpha}}$, we rewrite \eqref{eq:LTS(0)=double_sum} as
\begin{equation}\label{eq:LTS(0)=sum}
    \int_{0}^{\infty} \dd \tau\,
    e^{-\rho\tau}
    S(\tau\,\vert\,0)
    = 
    \frac{1}{1+\rho} + 
    \sum_{n=1}^{\infty} e^{-\frac{\rho+1}{\alpha}n} 
    \sum_{j=1}^{n}
    \frac{n^{n-j-1}}{(n-j)!}
    \frac{j\,\alpha^{j-n}}{(1+\rho)^{j+1}}.  
\end{equation}
For each term in the sum the Laplace transform can be inverted explicitly. The only singularity in the $\rho$-plane is a pole at $\rho=-1$. Computing the residue yields the inversion formula 
\begin{equation}\label{eq:LT_term_S(0)=}
    \mathcal{L}^{-1}_{\rho\mapsto\tau}
    \left[ 
        \frac{e^{-\frac{\rho+1}{\alpha}n}}
         {(\rho+1)^{j+1}}
    \right]
        = 
        \frac{(\alpha\tau - n)^j}{j!\, \alpha^j}
        \theta\left(\alpha \tau - n \right) e^{-\tau}.
\end{equation}
Applying then \eqref{eq:LT_term_S(0)=} to \eqref{eq:LTS(0)=sum} after some simplification we arrive at
\begin{equation}
    S(\tau\,\vert\,0) = 
    e^{-\tau} + 
    e^{-\tau} 
    \sum_{n=1}^{\infty}
    \theta(\alpha\tau-n)
    \sum_{j=1}^{n}
    \frac{n^{n-j-1}}{\alpha^n}
    \frac{(\alpha\tau-n)^j}{(j-1)! (n-j)!}.
\end{equation}
The sum over $j$ can be computed directly by recognizing the binomial expansion,
\begin{align}
    \nonumber
    \sum_{j=1}^{n}
    \frac{n^{n-j-1}}{\alpha^n}
    \frac{(\alpha\tau-n)^j}{(j-1)! (n-j)!}
    & = \frac{\alpha\tau-n}{\alpha^n \, n!}
    \sum_{j=0}^{n-1} 
    \binom{n-1}{j}n^{n-1-j} (\alpha\tau-n)^j
    \\
    & =
    \frac{\alpha \tau - n}{\alpha^n n!} (\alpha \tau)^{n-1}.
\end{align}
Therefore, after including the $n=0$ term, we obtain
\begin{equation}\label{eq:S(tau|0)=the_same_result}
    S(\tau\,\vert\,0) = 
    e^{-\tau} 
    \sum_{n=0}^{\infty}
    \theta(\alpha\tau-n)
    \frac{(\alpha\tau-n)}{\alpha n!} \tau^{n-1}.
\end{equation}
Due to the Heaviside function, the sum can be truncated. The upper limit is then replaced by $\lfloor \alpha \tau \rfloor$, and hence \eqref{eq:S(tau|0)=the_same_result} is exactly the representation obtained from the time domain approach \eqref{eq:S(t|0)=result_time_domain}. This completes the verification that both methods yield identical results, despite their radically different forms.

\subsubsection{Conditional moments}
\par Having established the equivalence of both approaches, we now exploit the Laplace-domain representation \eqref{eq:hatQ(rho,0)=} to compute conditional moments \eqref{eq:Ec[t^k]=def}. The key advantage of this approach is that moments can be extracted via a simple Taylor expansion, avoiding the nontrivial integrations required in the time domain. Below we compute the first two conditional moments. 

\par Recall that the conditional moments are given by
\begin{equation}
    \mathbb{E}_c\left[\,\tau^{k}\,\vert\,\beta\,\right] 
    = \frac{1}{1-S_\infty(\beta)}\int_{0}^{\infty}\dd\tau\, \tau^{k}\; \mathbb{P}\left[\,\tau\,\vert\,\beta\,\right],
\end{equation}
hence we can find them by expanding $Q(\rho\,\vert\,\beta)$ in series with respect to $\rho$. Specifically,
\begin{equation}\label{eq:Q(rho|0)=moment series}
    Q(\rho\,\vert\,\beta) = \big( 1 - S_{\infty}(\beta)\big)
    \sum_{k=0}^\infty \frac{(-\rho)^k}{k!} \mathbb{E}_{c}\left[\,\tau^k\,\vert\,\beta\,\right].
\end{equation}
Since we have an explicit representation for $Q(\rho\,\vert\,\beta)$ for the zero offset case $\beta=0$ in \eqref{eq:hatQ(rho,0)=}, we can directly expand it in a Taylor series around $\rho = 0$. The coefficients of this expansion yield both the survival probability $S_\infty(0)$ (from the zeroth-order term) and the conditional moments (from higher-order terms). The main technical task is constructing the expansion of the Lambert function. This can be accomplished using the differentiation rule (see Appendix~\ref{sec:app_LambertW}):
\begin{equation}\label{eq:d/dz W_0=rule}
    \dv{}{z} \Big[ W_0(z) \Big] = \frac{W_0(z)}{z(1 + W_0(z))},\qquad 
    z\notin \left\{-\frac{1}{e},0\right\},
\end{equation}
which implies
\begin{equation}\label{eq:W=expansion a<1}
    W_0\left[-\frac{1}{\alpha} e^{-\frac{\rho+1}{\alpha}}\right]
    \underset{\rho\to0}{=} 
    W_{0}\left[-\frac{1}{\alpha}e^{-\frac{1}{\alpha}}\right]
    % \\
    - \frac{W_{0}\left[-\frac{1}{\alpha}e^{-\frac{1}{\alpha}}\right]}{1+W_{0}\left[-\frac{1}{\alpha}e^{-\frac{1}{\alpha}}\right]}
    \frac{\rho}{\alpha}
    + O\left(\rho^2\right).
\end{equation}
The subtlety here is that for $z\ge-1$ the principal branch of the Lambert function satisfies $W_0(ze^{z})=z$ and hence 
\begin{equation}\label{eq:W_0(1/a e^1/a)=1/a}
    \alpha\ge 1:\qquad W_0\left[-\frac{1}{\alpha} e^{-\frac{1}{\alpha}}\right] = -\frac{1}{\alpha}.
\end{equation}
This identity has an important consequence: substituting it into \eqref{eq:hatQ(rho,0)=} shows that the denominator simplifies to $\rho + O(\rho^2)$ for $\alpha \ge 1$. This cancellation means the first-order expansion is insufficient, and we must construct higher-order terms to extract the moments. The direct computation using \eqref{eq:d/dz W_0=rule} and \eqref{eq:W_0(1/a e^1/a)=1/a} yields
\begin{multline}\label{eq:W=expansion a>1}
    \alpha>1:\quad 
    W_0\left[-\frac{1}{\alpha} e^{-\frac{\rho+1}{\alpha}}\right]
    \underset{\rho\to0}{=} -\frac{1}{\alpha} + \frac{1}{\alpha(\alpha-1)} \rho
    \\
    - \frac{1}{2(\alpha-1)^3} \rho^2 
    + \frac{(\alpha+2)}{6(\alpha-1)^5} \rho^3 + O\left(\rho^4\right).
\end{multline}
The rest involves straightforward algebraic manipulations. Substituting the expansion \eqref{eq:W=expansion a>1} into \eqref{eq:hatQ(rho,0)=}, we find that in the supercritical regime ($\alpha>1$)
\begin{equation}\label{eq:Q(rho,0)= a>1 expansion}
    \alpha>1:\quad Q(\rho\,\vert\, 0) \underset{\rho\to0}{=} 
    \frac{1}{\alpha} 
    - \frac{1}{2\alpha} \frac{1}{\alpha-1} \rho 
    +
    \frac{1}{12\alpha} \frac{2\alpha+1}{(\alpha-1)^3} \rho^2
    +O\left(\rho^3\right).
\end{equation}
Comparing \eqref{eq:Q(rho,0)= a>1 expansion} with \eqref{eq:Q(rho|0)=moment series} and equating coefficients, we obtain the survival probability at infinite time $S_\infty(0)$:
\begin{equation}
    \alpha>1:\quad
    S_\infty(0) = 1 - \frac{1}{\alpha},
\end{equation}
as well as the first two conditional  moments,
\begin{equation}\label{eq:first_two_moments=a>1}
    \alpha>1:\quad
    \mathbb{E}_c\left[\,\tau\,\vert\,0\,\right] = 
     \frac{1}{2} \frac{1}{\alpha-1},
    \qquad
    \mathbb{E}_c\left[\,\tau^2\,\vert\,0\,\right] = 
     \frac{1}{6} \frac{2\alpha+1}{(\alpha-1)^3}.
\end{equation}
Note that both moments diverge as $\alpha \to 1^+$. Moreover, $S_\infty(0) \to 1$ as $\alpha \to \infty$, which is indeed expected: when the boundary moves much faster than the process, it outpaces the trajectory and crossings become effectively impossible.

\par 
The subcritical case ($\alpha < 1$) proceeds similarly. Here, the identity \eqref{eq:W_0(1/a e^1/a)=1/a} does not apply, so the denominator in \eqref{eq:hatQ(rho,0)=} does not simplify, and the first-order expansion \eqref{eq:W=expansion a<1} suffices. Substituting it into \eqref{eq:hatQ(rho,0)=} yields:
\begin{multline}\label{eq:Q(rho,0)= a<1 expansion}
    \alpha<1:\quad Q(\rho\,\vert\,0) \underset{\rho\to0}{=} 
    1
    - \frac{1}{1+\alpha W_0\left[-\frac{1}{\alpha} e^{-\frac{1}{\alpha}}\right]} \rho
    \\
    +
    \frac{1}{1+ W_0\left[-\frac{1}{\alpha} e^{-\frac{1}{\alpha}}\right]}
    \frac{1}{\left(1+\alpha W_0\left[-\frac{1}{\alpha} e^{-\frac{1}{\alpha}}\right]\right)^2} \rho^2
    +O\left(\rho^3\right).
\end{multline}
The zeroth-order coefficient is $1$, confirming that $S_\infty(0)=0$: in the subcritical regime $\alpha<1$, the Poisson process (which grows at unit rate on average) outpaces the boundary (which moves at velocity $\alpha$), so the crossing occurs almost surely. The first two conditional moments read:
\begin{equation}\label{eq:first_moment=a<1}
   \alpha<1 : \quad  
   \mathbb{E}_c\left[\,\tau\,\vert\,0\,\right] = \frac{1}{1+\alpha W_0\left[-\frac{1}{\alpha} e^{-\frac{1}{\alpha}}\right]},
\end{equation}
and 
\begin{equation}\label{eq:second_moment=a<1}
    \alpha<1 : \quad \mathbb{E}_c\left[\,\tau^2\,\vert\,0\,\right] =
    \frac{1}{1+ W_0\left[-\frac{1}{\alpha} e^{-\frac{1}{\alpha}}\right]}
    \frac{2}{\left(1+\alpha W_0\left[-\frac{1}{\alpha} e^{-\frac{1}{\alpha}}\right]\right)^2}.
\end{equation}
As in the supercritical case, the moments diverge as $\alpha\to1^{-}$, which can be seen from~\eqref{eq:W_0(1/a e^1/a)=1/a}.

\par 
We emphasize that attempting to obtain closed representations for the moments from the direct time-domain approach would require evaluating quite involved integrals. For example, the mean first-passage time is given by
\begin{equation}\label{eq:int moment = explicit}
    \mathbb{E}_c\left[\tau\,\vert\,0\,\right] = 
    \frac{1}{1-S_\infty(0)}
    \int_{0}^{\infty} \dd\tau\, e^{-\tau}
    \frac{1}{\alpha}   
    \frac{\tau^{\lfloor\alpha \tau\rfloor}}{\lfloor\alpha \tau\rfloor!}
    (\alpha \tau - \lfloor\alpha \tau\rfloor) .
\end{equation} 
Not only is this integral nontrivial to compute, more importantly, it does not suggest that a simple closed-form representation like \eqref{eq:first_two_moments=a>1} exists at all.  At the same time, if evaluated numerically \eqref{eq:int moment = explicit} agrees with \eqref{eq:first_moment=a<1} for $\alpha<1$ and with \eqref{eq:first_two_moments=a>1} for $\alpha>1$.

\par Once the result is known, it can be verified as follows: first, represent the integral as an infinite sum with terms corresponding to segments with different integer values of 
$\lfloor \alpha\tau \rfloor$; second, after rescaling and shifting the integration variable, transform all integrals to the unit interval $[0,1]$; third, apply binomial expansions to recognize in the integrand expressions similar to \eqref{eq:W0^j=expansion}, but involving derivatives of the Lambert function; finally, compute the integrals to recover \eqref{eq:first_moment=a<1} and \eqref{eq:second_moment=a<1}. 

\par This verification is technically involved, and we therefore omit it here. What is crucial is that it requires knowing the answer beforehand; discovering \eqref{eq:first_two_moments=a>1} and \eqref{eq:first_moment=a<1} directly from \eqref{eq:int moment = explicit} would be exceedingly difficult without the Laplace-domain guidance. Expansions for the derivatives of the Lambert function in the integrand are simply not something one would think to look for without already expecting them to be there.

\subsubsection{Critical behavior of the conditional moments}
\par We now examine the behavior of the conditional moments as $\alpha$ approaches the critical value $\alpha\to1$. The explicit expressions \eqref{eq:first_two_moments=a>1}, \eqref{eq:first_moment=a<1}, and \eqref{eq:second_moment=a<1} suggest that the moments diverge as $(\alpha-1)^{1-2k}$. To quantify this divergence more precisely,  we introduce the function 
\begin{equation}\label{eq:F(R,alpha|beta)=def}
    \mathcal{F}(R,\alpha\,\vert\,\beta) 
    =
    \frac{1}{\left|\alpha-1\right|}
    \left[
        \frac{Q\left(R(\alpha-1)^2\,\vert\,\beta\right)}
             {1-S_\infty(\beta)}
        - 1
    \right].
\end{equation}
Recall that the conditional moments can be extracted from $Q(\rho\,\vert\,\beta)$ as in \eqref{eq:Q(rho|0)=moment series}. Substituting this expansion into \eqref{eq:F(R,alpha|beta)=def}, we immediately see that $\mathcal{F}(R,\alpha\,\vert\,\beta)$ is nothing but the generating function for the conditional moments:
\begin{equation}\label{eq:F(R,alpha|beta)=sum_moments}
    \mathcal{F}(R,\alpha\,\vert\,\beta) 
    =
    \sum_{k=1}^{\infty}
    \frac{\left|\alpha-1\right|^{2k-1}}{k!}
    \mathbb{E}_{c}\left[\,\tau^k\,\vert\,\beta\,\right]
    \left(-R\right)^{k}.
\end{equation}
The divergence of the conditional moments can now be obtained from the limiting behavior of $\mathcal{F}(R,\alpha\,\vert\,0)$ as $\alpha\to1$. This can be computed using the explicit form \eqref{eq:hatQ(rho,0)=} of $Q(\rho\,\vert\,0)$. A subtlety here is that $S_\infty(0)=0$ for $\alpha<1$ and $S_\infty(0)=1-\frac{1}{\alpha}$ for $\alpha>1$, so when taking the limit we must treat $\alpha\to1^+$ and $\alpha\to1^-$ separately. An explicit calculation shows that the two limits indeed coincide and yield
\begin{equation}\label{eq:F(R,a)=explicit}
     \lim_{\alpha\to1^{+}} \mathcal{F}(R,\alpha\,\vert\,0) = 
     \lim_{\alpha\to1^{-}} \mathcal{F}(R,\alpha\,\vert\,0) = 
     \frac{1}{2}\left(1 - \sqrt{1+2R}\right).
\end{equation}
Expanding \eqref{eq:F(R,a)=explicit} as a series in $R$ and comparing with \eqref{eq:F(R,alpha|beta)=sum_moments}, we find the asymptotic behavior of the moments  to be
\begin{equation}\label{eq:Ec=divergence_exact}
    \mathbb{E}_c\left[\,\tau^k\,\vert\,0\,\right]
    \underset{\alpha\to1}{\sim}
        \frac{(2k)!}{2^{k+1}\, k!\, (2k-1)} \left|\frac{1}{\alpha-1}\right|^{2k-1}.
\end{equation}
Interestingly, this reveals that both supercritical and subcritical approaches to the critical point exhibit identical singular behavior.

\subsection{Large offset}
\par We now turn to the opposite extreme and consider the large offset limit $\beta\to\infty$. This case provides a complementary perspective to the zero offset analysis and further demonstrates the power of the Laplace-domain approach, where the calculations become remarkably straightforward.

\subsubsection{Survival probability at infinite time}\label{subsec:Sinfty(beta->infty)}
\par The key quantity governing the behavior in this limit is the survival probability $S_\infty(\beta)$. This probability appears as an overall factor in the conditional moments \eqref{eq:Ec[t^k]=def}, so understanding its large-$\beta$ behavior is essential before analyzing the moments themselves. According to \eqref{eq:hat(Q)=def} and \eqref{eq:Sinfty=def} we have
\begin{equation}\label{eq:lim hatQ = int 1-S}
    \lim_{\rho\to0}\hat{Q}(\rho\,\vert\,\lambda) = \int_{0}^{\infty} \dd \beta\,
         e^{-\lambda\beta}
    \int_{0}^{\infty} \dd \tau\, \mathbb{P}\left[\,\tau\,\vert\,\beta\,\right]
    = \int_{0}^{\infty}\dd\beta\, e^{-\lambda\beta}
    \left(1 - S_{\infty}(\beta)\right).
\end{equation}
This means that the Laplace transform of $S_\infty(\beta)$ is given by 
\begin{equation}\label{eq:(Sinfty)LT=def}
    \hat{S}_{\infty}(\lambda) \equiv \int_{0}^{\infty}\dd\beta\, e^{-\lambda\beta}
    S_\infty(\beta) = 
    \frac{1}{\lambda} - \lim_{\rho\to0}\hat{Q}(\rho\,\vert\,\lambda).
\end{equation}
Therefore $\hat{S}_\infty(\lambda)$  can be found by taking the limit $\rho\to0$ in the explicit representation of $\hat{Q}(\rho\,\vert\,\lambda)$ given in~\eqref{eq:hat(Q)=explicit}:
\begin{equation}\label{eq:hat(Q)=explicit_1}
    \hat{Q}(\rho\,\vert\,\lambda) = 
    \frac{1}{\lambda} -
    \frac{1}{\lambda}
    \frac{\rho}{\rho+1-\alpha\lambda - e^{-\lambda}}
    \left(
        1 - \frac{\alpha \lambda}
                 {\rho+1 + \alpha W_0\left[
                     -\frac{1}{\alpha} e^{-\frac{\rho+1}{\alpha}}
                 \right] }
    \right).
\end{equation}
To take the limit in \eqref{eq:hat(Q)=explicit_1}, the key step is evaluating the denominator containing the Lambert function. This is easily done by utilizing the expansions we already obtained when computing the conditional moments for $\beta=0$. Specifically, from \eqref{eq:W=expansion a<1} we have
\begin{equation}\label{eq:denominator=a<1}
    \alpha<1 : \qquad \rho+1+\alpha W_0\left[-\frac{1}{\alpha} e^{-\frac{\rho+1}{\alpha}}\right]
    \underset{\rho\to0}{=} 
    1  + \alpha W_0 \left[-\frac{1}{\alpha} e^{-\frac{1}{\alpha}}\right] + O(\rho),
\end{equation}
and \eqref{eq:W=expansion a>1} implies that
\begin{equation}\label{eq:denominator=a>1}
    \alpha>1 : \qquad \rho+1+\alpha W_0\left[-\frac{1}{\alpha} e^{-\frac{\rho+1}{\alpha}}\right]
        \underset{\rho\to0}{=} \frac{\alpha}{\alpha-1} \, \rho+O\left(\rho^2\right).
\end{equation}
The denominator exhibits different behavior in the $\alpha>1$ and $\alpha<1$ cases. Substituting expansions \eqref{eq:denominator=a<1} and \eqref{eq:denominator=a>1} into \eqref{eq:hat(Q)=explicit_1} and taking the $\rho\to0$ limit, we find that the Laplace transform of $S_\infty(\beta)$ is given by
\begin{equation}\label{eq:(1-Sinfty)=LT}
    \hat{S}_\infty(\lambda) = 
    \frac{1}{\lambda} - \lim_{\rho\to0}\hat{Q}(\rho\,\vert\,\lambda)
    =
    \left\{
    \begin{aligned}
        &0, \qquad \alpha<1, \\
        &\frac{\alpha - 1}{\alpha\lambda - 1 + e^{-\lambda}}, \qquad \alpha>1.
    \end{aligned}\right.
\end{equation}
For the subcritical case $\alpha<1$, we have $S_\infty(\beta)=0$ for all $\beta$. This result is expected: when the boundary growth is slow enough, the process eventually crosses it with certainty regardless of the initial offset. For the supercritical case $\alpha>1$, the situation is different. There exists a finite survival probability, which we now analyze.

\begin{figure}[ht]
\includegraphics{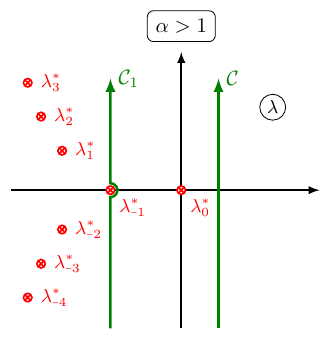}
\caption{Analytic structure of $\hat{S}_\infty(\lambda)$ in the $\lambda$-plane showing simple poles at $\lambda_\ell^*$ given by \eqref{eq:lambda_l(0)=}. The principal branch gives a pole at the origin, whose residue yields the limiting value $\lim_{\beta\to\infty} S_\infty(\beta) = 1$. The contour $\mathcal{C}$ in \eqref{eq:Sinfty=InverseLT_formally} is shifted to $\mathcal{C}_1$, picking up the residue at the origin. The remaining integral along $\mathcal{C}_1$ is dominated by the pole at $\lambda_{-1}^*$, which gives the subleading exponential correction \eqref{eq:Sinfty~e^-beta}.
}
\label{fig:Sinf(lambda)_anal}
\end{figure}

\par Formally inverting the Laplace transform \eqref{eq:(1-Sinfty)=LT} yields
\begin{equation}\label{eq:Sinfty=InverseLT_formally}
    S_{\infty}(\beta) = \frac{1}{2\pi\ii} \int_{\mathcal{C}} \dd \lambda\, e^{\beta \lambda} 
    \hat{S}_\infty(\lambda),
\end{equation}
where $\mathcal{C}$ is a vertical contour in the $\lambda$-plane such that all singularities lie to its left. To find the large-$\beta$ behavior of $S_\infty(\beta)$, we need to analyze the analytic structure of $\hat{S}_\infty(\lambda)$. This structure is rather straightforward: all singularities are simple poles arising from zeros of the denominator $\alpha\lambda - 1 + e^{-\lambda}$. These occur when
\begin{equation}\label{eq:lambda(0)_def_equation}
    \alpha\lambda - 1 + e^{-\lambda} = 0,
\end{equation}
or, equivalently,
\begin{equation}
    \left(\lambda-\frac{1}{\alpha}\right) e^{\lambda-\frac{1}{\alpha}} = -\frac{1}{\alpha} e^{-\frac{1}{\alpha}}.
\end{equation}
This is precisely the transcendental equation that defines the Lambert function (see Appendix~\ref{sec:app_LambertW}), hence the poles of $\hat{S}_\infty(\lambda)$ are located at the points $\lambda^*_\ell$ given by
\begin{equation}\label{eq:lambda_l(0)=}
    \lambda^*_\ell  = \frac{1}{\alpha}
    + W_{\ell}\left[-\frac{1}{\alpha}
    e^{-\frac{1}{\alpha}}\right], \qquad \ell\in\mathbb{Z},
\end{equation}
where $W_\ell(z)$ denotes the $\ell$-th branch of the Lambert function (see Fig.~\ref{fig:Sinf(lambda)_anal}). Note that due to \eqref{eq:W_0(1/a e^1/a)=1/a}, the solution corresponding to the principal branch is located at the origin, $\lambda_0^*=0$. This is the rightmost pole, and its residue is
\begin{equation}
    \Res_{\lambda=0}\left[ \frac{\alpha-1}{\alpha\lambda-1+e^{-\lambda}} \right] = 1,
\end{equation}
which implies
\begin{equation}\label{eq:limSinfty(beta)=1}
    \alpha>1:\qquad \lim_{\beta\to\infty}S_{\infty}(\beta) = 1.
\end{equation}
This is physically natural. If $\alpha>1$, the distance between the boundary and the Poisson process on average grows, so the larger the offset, the less likely it is for the process to reach the boundary. The interesting question here is how this asymptotic value is approached. This is governed by the singularity closest to the origin, which in our case is given by $\lambda_{-1}^{*}$. Indeed, shifting the contour $\mathcal{C}\mapsto\mathcal{C}_1$ as shown in Fig.~\ref{fig:Sinf(lambda)_anal}, we obtain
\begin{equation}\label{eq:Sinfty=InverseLT_formally_C1}
    S_{\infty}(\beta) = 1 + \frac{1}{2\pi\ii} \int_{\mathcal{C}_1} \dd \lambda\, e^{\beta \lambda} 
    \hat{S}_\infty(\lambda).
\end{equation}
Computing the integral via saddle-point approximation and using the explicit form of $\lambda_{-1}^{*}$ from \eqref{eq:lambda_l(0)=}, we find
\begin{equation}\label{eq:Sinfty~e^-beta}
    \alpha>1: \quad 
    1 - S_\infty(\beta) \underset{\beta\to\infty}{\asymp}
    \exp\left[
        \beta\left(
            \frac{1}{\alpha}
            + W_{-1}\left[-\frac{1}{\alpha}
            e^{-\frac{1}{\alpha}}\right]
        \right)
    \right].
\end{equation}

\par Note that since all singularities of $\hat{S}_\infty(\lambda)$ are simple poles, the inversion \eqref{eq:Sinfty=InverseLT_formally} can be performed via standard residue calculus techniques by summing over the residues at the points \eqref{eq:lambda_l(0)=}. This yields a representation of $S_\infty(\beta)$ in the form of an infinite sum, which we believe to be new:
\begin{equation}\label{eq:Sinf(beta)=sum_direct_Laplace}
    \alpha>1:\qquad  S_\infty(\beta) = \left(1-\frac{1}{\alpha}\right) 
    \sum_{\ell=-\infty}^{\infty} 
    \frac{ 
    \exp\left[
        \beta \left(\frac{1}{\alpha} + W_\ell\left[-\frac{1}{\alpha} e^{-\frac{1}{\alpha}} \right] \right)
    \right]
    }
    {1 + W_\ell\left[-\frac{1}{\alpha} e^{-\frac{1}{\alpha}} \right]}.
\end{equation}
This result is very different from the representation \eqref{eq:Sinfty(beta)=a>1res} obtained using the direct time-domain approach in \cite{P-59}. We establish their equivalence in Section~\ref{sec:general_offset}. For now, we simply observe that comparing \eqref{eq:Sinf(beta)=sum_direct_Laplace} with \eqref{eq:Sinfty(beta)=a>1res} yields a nontrivial identity for the Lambert function:
\begin{equation}\label{eq:sum_over_branches_identity}
    \sum_{\ell=-\infty}^{\infty} 
        \frac{e^{\beta\, W_\ell(z)} }
             {1 + W_\ell(z)}
    = 
    \sum_{j=0}^{\left\lfloor \beta \right\rfloor}
    \frac{(\beta - j)^j}{j!} z^j,
    \qquad
    z\in\left(-\frac{1}{e}, 0\right), 
    \qquad \beta>0.
\end{equation}
We emphasize that the representation \eqref{eq:Sinf(beta)=sum_direct_Laplace} is valid for $\beta>0$. A subtle technical point, which we mention but do not elaborate on, is that the limit $\beta\to0^+$ of the left-hand side in \eqref{eq:sum_over_branches_identity} differs from simply setting $\beta=0$ in each term.

\par 
The exponential decay \eqref{eq:Sinfty~e^-beta} has an important consequence: for $\alpha>1$, the conditional moments decay exponentially with $\beta$, making them less natural quantities to study in the large offset limit. Physically, this reflects that almost all trajectories starting far away fail to catch the fast-moving boundary, so conditioning on the rare crossing event becomes somewhat artificial. For the remainder of this section we thus focus on the subcritical regime, where crossing is certain and the conditional moments remain well-defined and physically meaningful as $\beta\to\infty$.

\subsubsection{Conditional moments in the subcritical regime} 
The conditional moments~\eqref{eq:Ec[t^k]=def} can be extracted directly from the Laplace-domain representation by series expansion of $\hat{Q}(\rho\,\vert\,\lambda)$ as shown in \eqref{eq:Q(rho|0)=moment series}. For the mean first-passage time, we have
\begin{equation}
    \int_{0}^{\infty} \dd\beta\, e^{-\lambda\beta}\,\big(1-S_\infty(\beta)\big)
    \mathbb{E}_{c}\left[\,\tau\,\vert\,\beta\,\right]  = 
    - \left. \pdv{}{\rho} \hat{Q}(\rho\,\vert\,\lambda) \right|_{\rho=0}.
\end{equation}
Since in the subcritical regime ($\alpha<1$) we have $S_\infty(\beta)=0$, using the explicit form \eqref{eq:hat(Q)=explicit} of $\hat{Q}(\rho\,\vert\,\lambda)$, we find the Laplace transform of the mean first-passage time:
\begin{equation}\label{eq:Ec[tau]=subctitical}
    \int_{0}^{\infty} \dd\beta\, e^{-\lambda\beta}\,
    \mathbb{E}_{c}\left[\,\tau\,\vert\,\beta\,\right] 
    = 
    -
    \frac{1}{\alpha\lambda-1+e^{-\lambda}}
    \left( \frac{1}{\lambda}
    - \frac{\alpha}
    {1 + \alpha W_0\left[-\frac{1}{\alpha} e^{-\frac{1}{\alpha}}\right]}
    \right).
\end{equation}
The large-$\beta$ behavior is extracted from the small-$\lambda$ expansion of this Laplace transform. Expanding around $\lambda=0$ yields
\begin{multline}
    \int_{0}^{\infty} \dd\beta\, e^{-\lambda\beta}\,
    \mathbb{E}_{c}\left[\,\tau\,\vert\,\beta\,\right] 
    \underset{\lambda\to0}{=}
    \frac{1}{\lambda^2} \,
    \frac{1}{(1-\alpha)}
    \\
    + 
    \frac{1}{\lambda}
    \frac{1}{2(1-\alpha)^2}
    \left( 1 - \frac{2 (1-\alpha)\alpha}
                    {1 + \alpha W_0\left[-\frac{1}{\alpha} e^{-\frac{1}{\alpha}}\right]}
    \right)
    +O(1).
\end{multline}
The $1/\lambda^2$ pole indicates linear growth with $\beta$, while the $1/\lambda$ pole gives the subleading constant term. Inverting the Laplace transform, we obtain
\begin{equation}\label{eq:E_c[tau|beta]_asymp}
    \mathbb{E}_{c}\left[\,\tau\,\vert\,\beta\,\right] 
    \underset{\beta\to\infty}{\sim}
    \frac{\beta}{(1-\alpha)}
    + 
    \frac{1}{2(1-\alpha)^2}
    \left( 1 - \frac{2 (1-\alpha)\alpha}
                    {1 + \alpha W_0\left[-\frac{1}{\alpha} e^{-\frac{1}{\alpha}}\right]}
    \right).
\end{equation}
Repeating the same procedure for the second moment, we obtain an analogous expression for the variance:
\begin{multline}\label{eq:Var[tau|beta]_asymp}
    \mathrm{Var}_c\left[\,\tau\,\vert\,\beta\,\right]
    \underset{\beta\to\infty}{\sim}
    \frac{\beta}{(1-\alpha)^3} 
    + \frac{7 + 8\alpha}{12(1-\alpha)^4}
    + \frac{\alpha^2}{(1-\alpha)^2}
        \frac{1}
             {\left( 1 + \alpha W_0\left[-\frac{1}{\alpha} e^{-\frac{1}{\alpha}}\right]\right)^2}
    \\ +
    \frac{\alpha}{(1-\alpha)^3} 
    \left(
        \frac{2\alpha-1}
             {1 + \alpha W_0\left[-\frac{1}{\alpha} e^{-\frac{1}{\alpha}}\right]}
        -
        \frac{2}{1 + W_0\left[-\frac{1}{\alpha} e^{-\frac{1}{\alpha}}\right]}
    \right).
\end{multline}

\par
\begin{figure}[ht]
    \centering
    \includegraphics[width=\linewidth]{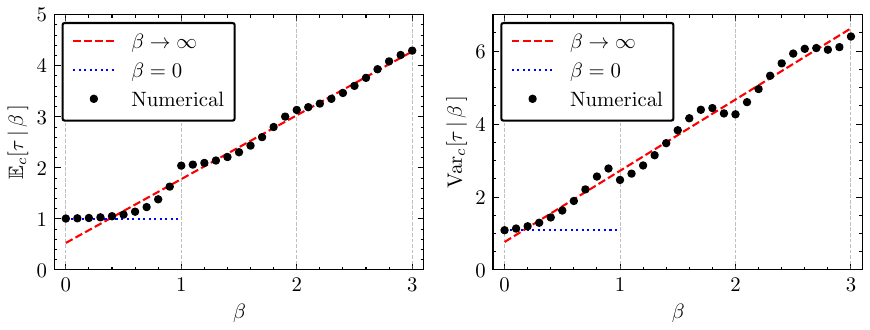}
    \caption{Comparison of analytic asymptotic predictions \eqref{eq:E_c[tau|beta]_asymp} and \eqref{eq:Var[tau|beta]_asymp} (red dashed lines) with numerical simulations (black circles) for the conditional mean and variance as functions of $\beta$ in the subcritical regime for $\alpha = 0{.}2$. 
    For comparison, we also show the mean and the variance for $\beta=0$, which are computed from the moments given in \eqref{eq:first_moment=a<1} and \eqref{eq:second_moment=a<1} (blue dotted lines).
    The numerical results are obtained by generating $10^7$ trajectories for each of $30$ equally spaced values of $\beta$ in the range $[0,3]$.
    }
    \label{fig:large_beta_comparison}
\end{figure}

 We conclude the analysis of conditional moments by comparing the analytic results \eqref{eq:E_c[tau|beta]_asymp} and \eqref{eq:Var[tau|beta]_asymp} against numerical simulations, shown in Fig.~\ref{fig:large_beta_comparison}.

\subsubsection{Large deviation form}
\par Both the mean and the variance of $\tau$ grow linearly with $\beta$. This suggests that the probability distribution admits a large deviation form (see \cite{T-09,T-18,BCKT-25}). Specifically, we expect that for $\tau\to\infty$ and $\beta\to\infty$ with fixed ratio $z = \alpha \tau/\beta$, the probability distribution becomes
\begin{equation}\label{eq:P=LDF_ansatz}
    \mathbb{P}\left[\,\tau\,\vert\,\beta\,\right] 
    \underset{\beta\to\infty}{\asymp}
    e^{ - \beta \, \Phi(z) },
    \qquad
    z = \frac{\alpha \tau}{\beta}.
\end{equation}
The rate function $\Phi(z)$ can be determined by exploiting two complementary perspectives on the Laplace transform.

\par First, by substituting the large deviation ansatz \eqref{eq:P=LDF_ansatz} into the definition of the Laplace transform,
\begin{equation}
    Q(\rho\,\vert\,\beta) =
    \int_{0}^{\infty} \dd\tau\,e^{-\rho\tau} \, \mathbb{P}\left[\,\tau\,\vert\,\beta\, \right]
    \underset{\beta\to\infty}{\asymp}
    \int_{0}^{\infty} \dd z\,e^{-\beta\left(\frac{\rho}{\alpha} z + \Phi(z)\right) },
\end{equation}
and then computing the integral in the saddle-point approximation, we obtain
\begin{equation}\label{eq:Q=LDF}
    Q(\rho\,\vert\,\beta)
    \underset{\beta\to\infty}{\asymp}
    \exp\left[ -\beta \min_{z}\left(\frac{\rho}{\alpha} z + \Phi(z) \right) \right] .   
\end{equation}
Alternatively, we can determine the $\beta\to\infty$ behavior of $Q(\rho\,\vert\,\beta)$ by locating the rightmost singularity of $\hat{Q}(\rho\,\vert\,\lambda)$ in the $\lambda$-plane. Recall that according to \eqref{eq:hat(Q)=explicit}, we have
\begin{equation}\label{eq:hatQ=explicit_LDF}
    \hat{Q}(\rho\,\vert\,\lambda) =
    \frac{1}{\lambda} -
    \frac{1}{\lambda}
    \frac{\rho}{\rho+1-\alpha\lambda - e^{-\lambda}}
    \left(
        1 - \frac{\lambda}
                 {\frac{\rho+1}{\alpha} + W_0\left[
                     -\frac{1}{\alpha} e^{-\frac{\rho+1}{\alpha}}
                 \right] }
    \right).       
\end{equation} 
At first glance, one expects a pole at $\lambda=0$ and singularities originating from the zeros of the denominator
\begin{equation}\label{eq:lambda(rho)_def_equation}
     \rho+1 - \alpha\lambda - e^{-\lambda} = 0,
\end{equation} 
This equation is essentially the same as \eqref{eq:lambda(0)_def_equation}, and the solutions are again given in terms of the Lambert function (recall \eqref{eq:lambda_l(0)=}):
\begin{equation}\label{eq:lambda*(rho)=def}
    \lambda_\ell^*(\rho) = \frac{\rho+1}{\alpha} + W_\ell\left[
        -\frac{1}{\alpha} e^{-\frac{\rho+1}{\alpha}}
    \right], \qquad
    \ell \in \mathbb{Z}.
\end{equation}
More careful analysis, however, reveals that the function is regular at $\lambda=0$ and at $\lambda^*_0(\rho)$. The rightmost singularity is thus given by $\lambda^{*}_{-1}(\rho)$, and therefore the $\beta\to\infty$ behavior is
\begin{equation}\label{eq:Q=singulary}
    Q(\rho\,\vert\,\beta) 
    \underset{\beta\to\infty}{\asymp}
    e^{ \beta \lambda_{-1}^*(\rho) }.
\end{equation}
Note that $\lambda^*_{-1}(\rho)$ is real and negative for $\rho\ge0$.

\par Comparing \eqref{eq:Q=singulary} with \eqref{eq:Q=LDF}, we immediately find that $\Phi(z)$ is the Legendre-Fenchel transform of $-\lambda^*_{-1}(\rho)$, i.e.,
\begin{equation}
    \min_{z}\left(\frac{\rho}{\alpha} z + \Phi(z) \right)
    = - \lambda_{-1}^{*}(\rho).
\end{equation}
Inverting this transform, we arrive at the parametric representation for the rate function~$\Phi(z)$:
\begin{equation}\label{eq:Phi(z)=parametric} 
    \Phi(z) = -
    \min_{\rho}\left(
        \frac{\rho}{\alpha} z
        +
        \frac{\rho+1}{\alpha}
        +
        W_{-1}\left[ - \frac{1}{\alpha} e^{-\frac{\rho+1}{\alpha}} \right]
    \right).
\end{equation}
The minimization can be performed explicitly. A property of the secondary real branch of the Lambert function, valid for $z\le-1$, is $W_{-1}(z e^{z})=z$. This implies that, similarly 
to \eqref{eq:W_0(1/a e^1/a)=1/a},
\begin{equation}\label{eq:W_-1(1/a e^1/a)=1/a}
    \alpha< 1:\qquad W_{-1}\left[-\frac{1}{\alpha} e^{-\frac{1}{\alpha}}\right] = -\frac{1}{\alpha}.
\end{equation}
Taking the derivative of \eqref{eq:Phi(z)=parametric} with respect to $\rho$ and using the differentiation rule \eqref{eq:d/dz W_0=rule}, we find that the minimum is reached at
\begin{equation}
    \rho^* = \alpha\left(1 + \frac{1}{z}\right) - \alpha \log\left[ \alpha \left(1+\frac{1}{z}\right) \right] - 1,
\end{equation}
which yields the explicit form of the rate function
\begin{equation}\label{eq:Phi(z)=explicit}
    \Phi(z) = \frac{z}{\alpha} - z-1 + (z+1)\log\left[\alpha\left(1+\frac{1}{z}\right)\right].
\end{equation}
Remarkably, the Lambert function does not appear in this final result.

\par To conclude the analysis, we extract some asymptotic properties of the probability distribution from the rate function. From \eqref{eq:Phi(z)=explicit}, it is clear that $\Phi(z)$ is a convex function with a minimum at $z^*=\alpha/(1-\alpha)$, and one can easily obtain its expansions in three different regimes:
\begin{equation}\label{eq:Phi(z)=asymptotics}
    \Phi(z) = 
    \left\{
    \begin{aligned}
        & - \log \frac{z}{\alpha} -1,
            \quad z\to0,\\
        & \frac{(1-\alpha)^3}{2\alpha^2}\left(z-z^*\right)^2 , 
            \quad z\to z^* = \frac{\alpha}{1-\alpha},\\
        & \left( \log\alpha + \frac{1}{\alpha} - 1\right) z,
            \quad z\to\infty.
    \end{aligned}
    \right.
\end{equation}
The behavior of $\Phi(z)$ near $z^*$ determines the form of $\mathbb{P}[\tau\,\vert\,\beta]$ near the typical crossing time. Specifically, from \eqref{eq:Phi(z)=asymptotics} we deduce that
\begin{equation}\label{eq:P~typical}
    \mathbb{P}\left[\,\tau\,\vert\,\beta\,\right]
    \underset{\beta\to\infty}{\sim}
    \exp\left[ - \frac{1}{2} 
            \frac{\left(\tau - \mathbb{E}_\text{as}[\,\tau\,]\right)^2}
                 {\mathrm{Var}_\text{as}\left[\,\tau\,\right]} \right],
    \qquad \tau\approx \mathbb{E}_\text{as}\left[\,\tau\,\right],
\end{equation}
which shows that the distribution is Gaussian near typical values, with asymptotic mean and variance
\begin{equation}\label{eq:E_inf=,Var_inf=}
    \mathbb{E}_\text{as}\left[\,\tau\,\right] = \frac{\beta}{1-\alpha},\qquad
    \mathrm{Var}_\text{as}\left[\,\tau\,\right] = \frac{\beta}{(1-\alpha)^3}.
\end{equation}
Both the mean and the variance are in agreement with the asymptotic expansions \eqref{eq:E_c[tau|beta]_asymp} and \eqref{eq:Var[tau|beta]_asymp}.

\par The left tail of the distribution, corresponding to atypically small values of $\tau$, is given by
\begin{equation}\label{eq:P~left_tail}
    \mathbb{P}\left[\,\tau\,\vert\,\beta\,\right] 
    \underset{\beta\to\infty}{\sim}
    \exp\left[\beta + \beta\log\frac{\tau}{\beta} \right],\qquad
    \tau \ll \mathbb{E}_\text{as}\left[\,\tau\,\right],
\end{equation}
and for the right tail we have
\begin{equation}\label{eq:P~right_tail}
    \mathbb{P}\left[\,\tau\,\vert\,\beta\,\right]
    \underset{\beta\to\infty}{\sim}
    \exp\Big[
        - \tau \left(1-\alpha+\alpha\log\alpha\right)
    \Big],
    \qquad 
    \tau \gg \mathbb{E}_\text{as}\left[\,\tau\,\right].
\end{equation}
These tail asymptotics reveal that large deviations from the mean first-passage time are exponentially suppressed, with the right tail (late crossings) decaying exponentially in $\tau$.

\par We should stress that the Laplace-domain approach yields the large deviation form in a straightforward manner, whereas deriving it directly from the time-domain representation 
\eqref{eq:P=result_time_domain_res} appears considerably more challenging, and we are not aware of a systematic way in which it can be done.

\section{General offset}\label{sec:general_offset}
\par Having analyzed two extreme cases and developed intuition about the first-passage properties, we now turn to the general offset case. In this section, we obtain the asymptotic behavior of the survival probability $S(\tau\,\vert\,\beta)$ as $\tau\to\infty$ and compute the conditional mean first-passage time for arbitrary $\beta$.

\subsection{Survival probability}\label{sec:S(tau|beta)-general}
\par We start with the survival probability
$S(\tau\,\vert\,\beta)$. Recall that its double Laplace
transform is given by \eqref{eq:hat(S)=explicit} and reads
\begin{equation}\label{eq:hatS=explicit_general}
    \hat{S}(\rho\,\vert\,\lambda)=
    \frac{1}{
        \lambda\left( 1 + \rho - \alpha\lambda - e^{-\lambda}\right)
    }
    \left(
    1 - \frac{\alpha \lambda}
             {\rho + 1 + \alpha W_0\left[ -\frac{1}{\alpha} e^{-{\frac{\rho+1}{\alpha}}}
    \right]} \right).
\end{equation}
To extract the asymptotic behavior as $\tau\to\infty$, we analyze the singularity structure of \eqref{eq:hatS=explicit_general} in the $\rho$-plane. 

\par There are several potential sources of singularities in \eqref{eq:hatS=explicit_general}. The first is the overall factor that might induce a simple pole at
\begin{equation}\label{eq:tilde(rho)=def}
    \tilde{\rho} = \alpha\lambda-1 + e^{-\lambda}.
\end{equation}
The second is the Lambert function. 
However, detailed analysis reveals that the apparent pole at \eqref{eq:tilde(rho)=def} does not contribute to the large-$\tau$ asymptotics. To see this, we note that
\begin{equation}
    W_0\left[ -\frac{1}{\alpha} e^{-{\frac{\tilde{\rho}+1}{\alpha}}}\right]
    =
    W_0\left[ 
        - \frac{1}{\alpha} e^{-\lambda} \left\{ 
            e^{ - \frac{1}{\alpha} e^{-\lambda} }
            \right\}
    \right].
\end{equation}
Recalling the property \eqref{eq:W_0(1/a e^1/a)=1/a} of the principal branch of the Lambert function, we immediately see that
\begin{equation}\label{eq:W0=-1/a e^-l_identity}
    \alpha e^\lambda \ge 1:\qquad 
    W_0\left[ -\frac{1}{\alpha} e^{-{\frac{\tilde{\rho}+1}{\alpha}}}\right]
    = - \frac{1}{\alpha} e^{-\lambda},
\end{equation}
and hence 
\begin{equation}
    \alpha e^\lambda \ge 1:\qquad 
    \tilde{\rho}+1+\alpha W_{0}\left[ -\frac{1}{\alpha} e^{-\frac{\tilde{\rho}+1}{\alpha}} \right] = \alpha\lambda.
\end{equation}
Therefore at $\rho=\tilde{\rho}$, the second factor in \eqref{eq:hatS=explicit_general} equals zero. This cancels the pole originating from the first factor, rendering the function regular at $\rho=\tilde{\rho}$.

\par A small subtlety here is that the arguments above hold for $\alpha e^{\lambda}\ge1$. However, this condition is not very restrictive, and we can lift it by first formally inverting the Laplace transform with respect to $\lambda$ similarly to \eqref{eq:Sinfty=InverseLT_formally} and then choosing the contour with sufficiently large $\lambda$. Since the identity \eqref{eq:W0=-1/a e^-l_identity} can actually be extended to the complex arguments (see Fig.~\ref{fig:LambertComplexStructure} in the Appendix~\ref{sec:app_LambertW}), the pole at $\tilde{\rho}$ has no effect on the large-$\tau$ behavior.

\par The other potential source of singularities of $\hat{S}(\rho\,\vert\,\lambda)$ in the $\rho$-plane comes from the second term in \eqref{eq:hatS=explicit_general}, which can exhibit a pole when the denominator vanishes and branch cuts from the Lambert function. First, we examine the possible pole. The denominator vanishes when
\begin{equation}\label{eq:Lambert_inducedPole}
    -\frac{\rho+1}{\alpha} = W_0\left[ -\frac{1}{\alpha} 
    e^{-{\frac{\rho+1}{\alpha}}}\right].
\end{equation}
Applying the definition of the Lambert function, $W(z)e^{W(z)} = z$, we obtain
\begin{equation}
    -\frac{\rho+1}{\alpha} e^{ - \frac{\rho+1}{\alpha} } = 
    -\frac{1}{\alpha} e^{-\frac{\rho+1}{\alpha}},
\end{equation}
which gives $\rho=0$ as the only candidate solution. Note that this algebraic step discards information about which branch of the Lambert function is involved. To verify whether $\rho=0$ is indeed a solution of \eqref{eq:Lambert_inducedPole}, we substitute it back and invoke \eqref{eq:W_0(1/a e^1/a)=1/a}. This identity holds for the principal branch $W_0$ only when $\alpha\ge1$; for $\alpha<1$, the value $\rho=0$ would instead correspond to $W_{-1}$ branch, as shown in \eqref{eq:W_-1(1/a e^1/a)=1/a}. We conclude that $\rho=0$ is a pole of $\hat{S}(\rho\,\vert\,\lambda)$ if and only if $\alpha\ge1$.

\par
\begin{figure}[ht]
    \centering
    \includegraphics{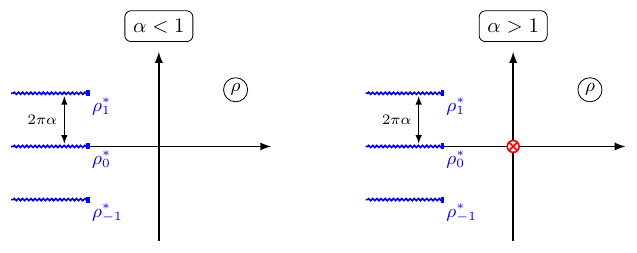}
    \caption{Schematic representation of the analytic structure of $\hat{S}(\rho\,\vert\,\lambda)$ in the $\rho$-plane. Branch cuts  emanate from the branch points \eqref{eq:rho^*=cuts}. For $\alpha>1$, there is an additional pole at $\rho=0$, which gives the survival probability at infinite time $\hat{S}_\infty(\lambda)$.}
    \label{fig:analytic_structure_S_in_rho}
\end{figure}

 Finally, the Lambert function has a branching point when its argument equals $-1/e$, which induces branch cuts in $\hat{S}(\rho\,\vert\, \lambda)$ with branch points at $\rho^*_\ell$ given by
\begin{equation}\label{eq:rho^*=cuts}
     \rho^*_\ell = \alpha-1 - \alpha\log\alpha +
     2\pi \ii \, \alpha \ell, \qquad \ell \in \mathbb{Z}.
\end{equation}
All these branch points have the same real part, namely $\Re[\rho^*_\ell] = \rho_0^*=\alpha-1-\alpha\log\alpha$. The schematic representation of the analytic structure of \eqref{eq:hatS=explicit_general} is shown in Fig.~\ref{fig:analytic_structure_S_in_rho}.

\par We now extract the asymptotic behavior. For the subcritical case $\alpha<1$, there is no pole at $\rho=0$, and the rightmost singularities are the branch points at $\rho^*_\ell$. Standard singularity analysis of Laplace transforms then yields
\begin{equation}\label{eq:S(lambda)_asymp_alpha<1}
    \alpha<1:\qquad
    \int_{0}^{\infty} \dd \beta \, e^{-\lambda\beta}\,
    S(\tau\,\vert\,\beta)
    \underset{\tau\to\infty}{\asymp}
    e^{\rho_0^* \, \tau}.
\end{equation}
Importantly, the exponential decay rate in $\tau$ is independent of $\beta$, and all information about the offset is contained in the multiplicative prefactor, which is subexponential as $\tau\to\infty$. Thus, \eqref{eq:S(lambda)_asymp_alpha<1} translates into
\begin{equation}\label{eq:S_asymp_alpha<1}
    \alpha<1:\qquad
    S(\tau\,\vert\,\beta)
    \underset{\tau\to\infty}{\asymp}
    \exp\Big[ -\tau (1-\alpha + \alpha\log\alpha) \Big],
\end{equation}
where we have used the explicit form of $\rho^*_0$ from \eqref{eq:rho^*=cuts}.

\par Note that the survival probability \eqref{eq:S_asymp_alpha<1} decays with the same rate as the right tail of $\mathbb{P}[\,\tau\,\vert\,\beta\,]$ computed in \eqref{eq:P~right_tail} in the limit $\beta\to\infty$. This provides a simple self-consistency check. Due to \eqref{eq:S(t|beta)=def}, for the subcritical regime we have
\begin{equation}\label{eq:S(tau|beta)=int_tau^infty P}
    \alpha<1:\qquad
    S(\tau\,\vert\,\beta) 
        = \int_{\tau}^{\infty} \dd \bar{\tau}\,\mathbb{P}\left[\,\bar{\tau}\,\vert\,\beta\,\right].
\end{equation}
For sufficiently large $\tau$, we can replace the probability distribution in \eqref{eq:S(tau|beta)=int_tau^infty P} by its asymptotic behavior \eqref{eq:P~right_tail}. Computing the integral then yields \eqref{eq:S_asymp_alpha<1}.

\par For the supercritical case $\alpha>1$, the pole at $\rho=0$ gives the survival probability at infinite time. Evaluating the residue, we find
\begin{equation}\label{eq:S_infty_alpha>1}
    \alpha>1:\qquad
    \lim_{\tau\to\infty}
    \int_{0}^{\infty} \dd\beta\, e^{-\lambda\beta}
        S(\tau\,\vert\,\beta)
    = 
    \Res_{\rho=0} \Big[  \hat{S}(\rho\,\vert\,\lambda) \Big] = 
    \frac{\alpha-1}{\alpha\lambda-1+e^{-\lambda}}.
\end{equation}
This is precisely the Laplace transform of the survival probability at infinite time from \eqref{eq:(1-Sinfty)=LT}. The decay toward this limiting value is again controlled by the branch points, i.e.,
\begin{equation}\label{eq:S_decay_alpha>1}
    \alpha>1:\qquad
    S(\tau\,\vert\,\beta) - S_{\infty}(\beta)
    \underset{\tau\to\infty}{\asymp}
    e^{ \rho^*_0 \, \tau }.
\end{equation}

\par Remarkably, the exponential decay rate $1-\alpha+\alpha\log\alpha$ is identical on both sides of the critical value $\alpha=1$, allowing us to write the result in a unified form:
\begin{equation}\label{eq:S_universal_decay}
    S(\tau\,\vert\,\beta) - S_{\infty}(\beta)
    \underset{\tau\to\infty}{\asymp}
    \exp\left[ - \frac{\tau}{\xi(\alpha)} \right],\qquad
    \xi(\alpha) = \frac{1}{1-\alpha + \alpha\log\alpha},
\end{equation}
where $\xi(\alpha)$ is the characteristic relaxation time. This is exactly the result stated in \eqref{eq:S-Sinf ~ e^}.

\par To conclude the analysis of the asymptotic behavior of the survival probability, we need to address two questions:  first, we must characterize the behavior at the critical point $\alpha=1$, since the corresponding time scale $\xi(\alpha)$ diverges as $\alpha\to1$ and \eqref{eq:S_universal_decay} is no longer valid; second, we must find the limit $S_\infty(\beta)$ from its Laplace transform \eqref{eq:S_infty_alpha>1}.

\paragraph{Critical case} 
If $\alpha=1$, then in the $\rho$-plane the pole at $\rho=0$ coincides with the branch point $\rho^*_0$, and hence the analysis of the analytic structure must be modified. Specifically, setting $\alpha=1$ in \eqref{eq:hatS=explicit_general} and expanding in series as $\rho\to0$ yields
\begin{equation}\label{eq:hatS= critical_expansion}
    \alpha=1:\qquad
    \hat{S}(\rho\,\vert\,\lambda)
    \underset{\rho\to0}{=} 
    \frac{1}{\sqrt{\rho}}\, 
        \frac{1}{\sqrt{2}} \frac{1}{\lambda+e^{-\lambda}-1} + O(1).
\end{equation}
This singular behavior suggests 
\begin{equation}\label{eq:S(tau|lambda)~critical}
    \alpha=1:\qquad
    \int_{0}^{\infty} \dd\beta\, 
        e^{-\lambda\beta} S(\tau\,\vert\,\beta)
    \underset{\tau\to\infty}{\sim}
    \frac{1}{\sqrt{2\pi\tau}}
    \frac{1}{\lambda+e^{-\lambda}-1}.
\end{equation}
We now invert the Laplace transform using essentially the same approach as in Section~\ref{subsec:equivalence} for the survival probability in the $\beta=0$ case. Specifically, we rely on the series representation
\begin{equation}\label{eq:critical_sum_identity}
    \frac{1}{\alpha\lambda-1+e^{-\lambda}}
    = 
    - \sum_{j=0}^\infty 
        \frac{e^{-j \lambda}}
             {(1-\alpha \lambda)^{j+1}},
\end{equation}
and the inversion formula for the Laplace transform similar to \eqref{eq:LT_term_S(0)=}, namely
\begin{equation}\label{eq:LT_term_l->beta=}
    \mathcal{L}^{-1}_{\lambda\mapsto\beta}\left[
        -\frac{e^{-j \lambda}}
             {(1-\alpha \lambda)^{j+1}}
    \right]
    =
    \frac{(-1)^{j}(\beta-j)^{j}}{j! \; \alpha^{j+1}} \theta(\beta-j) e^{\frac{\beta-j}{\alpha}}. 
\end{equation}
Substituting \eqref{eq:critical_sum_identity} into \eqref{eq:S(tau|lambda)~critical}, applying \eqref{eq:LT_term_l->beta=} term-by-term, and setting $\alpha=1$, we obtain
\begin{equation}\label{eq:S(tau|beta)~critical}
    \alpha=1:\qquad
    S(\tau\,\vert\,\beta)
    \underset{\tau\to\infty}{\sim}
    \frac{1}{\sqrt{2\pi\tau}}
    \sum_{j=0}^{\lfloor \beta \rfloor}
    \frac{(-1)^{j}}{j!}(\beta-j)^j e^{\beta-j}.
\end{equation}
In other words, at the critical point $\alpha=1$, the exponential decay \eqref{eq:S_universal_decay} is replaced by a much slower algebraic decay.

\begin{figure}[h]
    \includegraphics{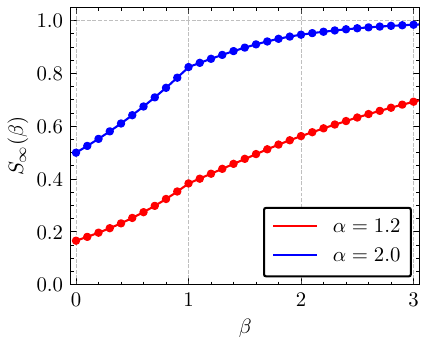}
    \caption{Survival probability at infinite time $S_\infty(\beta)$ as a function of the offset $\beta$ for  for $\alpha = 1{.}2$ (red) and $\alpha=2$ (blue). 
    Solid lines correspond to the analytical result~\eqref{eq:Sinfty(beta)=sum}, while circles correspond to direct numerical simulations obtained by generating $10^6$ independent trajectories of the Poisson process for each of 30 equally spaced values of $\beta$ in the range $[0,3]$. To approximately cater for the $\tau \to \infty$ limit, each trajectory was evolved until $1000$ jumps occurred.}\label{fig:Sinfty(beta)=numerical}
\end{figure}

\paragraph{Closed form representation for $S_\infty(\beta)$} 
Finally, we invert the Laplace transform \eqref{eq:S_infty_alpha>1}. Representing $\hat{S}_\infty(\lambda)$ as an infinite sum, we arrive at
\begin{equation}\label{eq:Sinfty=asymp}
    \alpha>1:\qquad \hat{S}_\infty(\lambda) = 
    (1-\alpha)\sum_{j=0}^{\infty} \frac{e^{-j \lambda}}{(1-\alpha\lambda)^{j+1}} .
\end{equation}
We then use \eqref{eq:LT_term_l->beta=} and  invert the Laplace transform term-by-term to obtain
\begin{equation}\label{eq:Sinfty(beta)=sum}
    \alpha>1:\qquad  S_\infty\left(\beta\right) = \left(1-\frac{1}{\alpha}\right)
        \sum_{j=0}^{ \lfloor \beta \rfloor } 
         \frac{1}{\alpha^j}\frac{(-1)^{j}}{j!}
        \left(\beta-j\right)^j 
        e^{\frac{\beta - j}{\alpha}} .
\end{equation}
Note that when computing \eqref{eq:S(tau|beta)~critical} and \eqref{eq:Sinfty(beta)=sum}, we used a somewhat unorthodox approach to the inversion of the Laplace transform. This was strongly motivated by the intuition gained from the time-domain approach and the calculations performed in Section~\ref{subsec:equivalence} to establish the equivalence of the two methods in the case $\beta=0$. This yielded simple closed-form representations that resemble \eqref{eq:S(t,beta)=result_time_domain}. For completeness, we provide the derivation from the time-domain approach in Appendix~\ref{sec:app_Sinfty}. The comparison of \eqref{eq:Sinfty(beta)=sum} with numerical simulations is shown in Fig.~\ref{fig:Sinfty(beta)=numerical}.

\subsection{Mean first-passage time}
\par Having computed the survival probability for arbitrary offset $\beta$, we proceed to the computation of the conditional mean first-passage time \eqref{eq:Ec[t^k]=def}. The calculations are slightly different in the cases $\alpha>1$ and $\alpha<1$, and therefore we treat them separately.

\paragraph{Subcritical regime}
We begin with the subcritical case $\alpha<1$, where $S_\infty(\beta)=0$. Recall that according to \eqref{eq:Ec[tau]=subctitical}, the Laplace transform of the mean first-passage time is given by
\begin{equation}\label{eq:Ec[tau]=subctitical_1}
    \int_{0}^{\infty} \dd\beta\, e^{-\lambda\beta}\,
    \mathbb{E}_{c}\left[\,\tau\,\vert\,\beta\,\right] 
    = 
    -\frac{1}{\alpha\lambda-1+e^{-\lambda}}
    \left( \frac{1}{\lambda}
    - \frac{\alpha}
    {1 + \alpha W_0\left[-\frac{1}{\alpha} e^{-\frac{1}{\alpha}}\right]}
    \right).
\end{equation}
Following the strategy employed for the survival probability in Section~\ref{subsec:equivalence}, we use the identity \eqref{eq:critical_sum_identity} to express the Laplace transform as an infinite sum:
\begin{equation}\label{eq:LT[Ec]=subcritical_expansion}
    \int_{0}^{\infty} \dd\beta\, e^{-\lambda\beta}\,
    \mathbb{E}_{c}\left[\,\tau\,\vert\,\beta\,\right] 
    = 
    \left( \frac{1}{\lambda}
    - \frac{\alpha}
    {1 + \alpha W_0\left[-\frac{1}{\alpha} e^{-\frac{1}{\alpha}}\right]}
    \right)
    \sum_{j=0}^{\infty} \frac{e^{-j\lambda}}{(1-\alpha\lambda)^{j+1}}.
\end{equation}
Each term in the sum can now be inverted by computing residues in the $\lambda$-plane. The key inversions are
\begin{align}
    \label{eq:L-1(() ) = simple expression}
    & \mathcal{L}^{-1}_{\lambda\to\beta}\left[
        \frac{e^{-j\lambda}}{(1-\alpha\lambda)^{j+1}}
    \right] =
    -
    \frac{(-1)^{j}}{j!}
    \frac{\left(\beta-j\right)^{j}}{\alpha^{j+1}}
    e^{\frac{\beta-j}{\alpha}}\;
    \theta\left(\beta-j\right),
    \\
    \label{eq:L-1(1/l () ) = incomplete_gamma}
    & \mathcal{L}^{-1}_{\lambda\to\beta}\left[
        \frac{1}{\lambda}\frac{e^{-j\lambda}}{(1-\alpha\lambda)^{j+1}}
    \right]
    =
    -\frac{(-1)^{j}}{j!}
    \theta\left(\beta-j\right)\;
    \int_{0}^{\frac{\beta-j}{\alpha}} \dd y\, y^{j} e^{y}. 
\end{align}
The first identity follows from the residue at $\lambda=1/\alpha$, while the second requires computing residues at both $\lambda=0$ and $\lambda=1/\alpha$. The latter gives rise to an incomplete gamma function, which appears as the integral in \eqref{eq:L-1(1/l () ) = incomplete_gamma}.

\par
\begin{figure}[ht]
    \centering
    \includegraphics{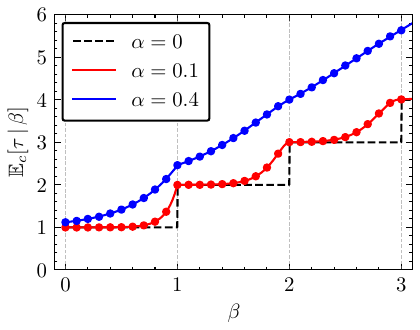}
    \caption{Conditional mean first-passage time $\mathbb{E}_c\left[\,\tau\,\vert\,\beta\,\right]$ as a function of the offset $\beta$ for $\alpha=0{.}1$ (red) and $\alpha=0{.}4$ (blue). Solid lines correspond to the analytical result \eqref{eq:Ec=subcritical_answer}, while circles correspond to direct numerical simulations obtained by generating $10^{6}$ independent trajectories of the Poisson process for each of $30$ equally spaced values of $\beta$ in the range $[0,3]$. The black dashed line shows the asymptotic result for $\alpha=0$ from \eqref{eq:Ec_alpha=0}.}
    \label{fig:Ec(a<1)}
\end{figure}

 Combining \eqref{eq:L-1(1/l () ) = incomplete_gamma} and \eqref{eq:L-1(() ) = simple expression} with \eqref{eq:LT[Ec]=subcritical_expansion}, we obtain the closed-form representation for the mean first-passage time in the subcritical regime:
\begin{equation}\label{eq:Ec=subcritical_answer}
    \alpha<1:\quad
    \mathbb{E}_{c}\left[\,\tau\,\vert\,\beta\,\right] 
    = 
    \sum_{j=0}^{\lfloor \beta \rfloor}
    \frac{(-1)^{j}}{\alpha^j\, j!}
    \left\{
    \frac{\left( \beta-j\right)^{j} e^{\frac{\beta-j}{\alpha}}}
         {1 + \alpha W_0\left[-\frac{1}{\alpha} e^{-\frac{1}{\alpha}}\right]}
    -\frac{1}{\alpha}
    \int_{0}^{\beta-j}\dd y\, y^{j} e^{\frac{y}{\alpha} }
    \right\}.
\end{equation}
While this expression involves both algebraic terms and integrals, it is exact and can be readily evaluated numerically for any value of $\beta$ (see Fig.~\ref{fig:Ec(a<1)}). As a consistency check, one can easily verify that setting $\beta=0$ reproduces \eqref{eq:first_moment=a<1}. 

\par Another interesting observation concerns the limiting case $\alpha=0$, in which the boundary becomes stationary and fixed at height $\beta$. In this limit, the process must accumulate exactly $\lfloor \beta \rfloor + 1$ jumps to cross the boundary. Taking the limit $\alpha\to0$ in \eqref{eq:Ec[tau]=subctitical_1} and inverting the Laplace transform yields
\begin{equation}\label{eq:Ec_alpha=0}
    \alpha=0:\quad 
    \mathbb{E}_{c}\left[\,\tau\,\vert\,\beta\,\right] = \lfloor \beta \rfloor + 1,
\end{equation}
which is exactly the expected time for a unit-rate Poisson process \eqref{eq:poisson_dist_unit} to perform $\lfloor \beta \rfloor + 1$ jumps. 
Moreover, in the limit $\alpha=0$ one can obtain the explicit representation for higher moments, 
\begin{equation}\label{eq:Ec[k]_alpha=0}
    \alpha=0:\quad 
    \mathbb{E}_{c}\left[\,\tau^k\,\vert\,\beta\,\right] = 
    \frac{(k + \lfloor \beta \rfloor)!}{\lfloor \beta \rfloor!}.
\end{equation}

\paragraph{Supercritical regime}
For the supercritical regime $\alpha>1$, the computation is similar but involves the prefactor $(1-S_\infty(\beta))$. The Laplace transform of the conditional mean first-passage time is given by
\begin{equation}
    \int_{0}^{\infty}\dd\beta \,e^{-\lambda\beta}  
    (1-S_\infty(\beta))\;
        \mathbb{E}_c\left[\,\tau\,\vert\,\beta\,\right]
    = 
    \left. - \pdv{}{\rho} \hat{Q}(\rho\,\vert\,\lambda) \right|_{\rho=0}.
\end{equation}
Evaluating the derivative using the explicit form \eqref{eq:hat(Q)=explicit}, the differentiation rule for the Lambert function \eqref{eq:d/dz W_0=rule}, and the identity \eqref{eq:W_0(1/a e^1/a)=1/a} for its principal branch, we arrive at
\begin{equation} 
    \int_{0}^{\infty}\dd\beta \,e^{-\lambda\beta}  
    (1-S_\infty(\beta))\;
        \mathbb{E}_c\left[\,\tau\,\vert\,\beta\,\right]
    = 
    \frac{\frac{1}{2}\frac{1}{\alpha-1} - \frac{1}{\lambda}}
           {\alpha\lambda-1+e^{-\lambda}}
    +
    \frac{\alpha-1}{\left(\alpha\lambda-1+e^{-\lambda}\right)^{2}}.
\end{equation}

\par
\begin{figure}[ht]
    \centering
    \includegraphics{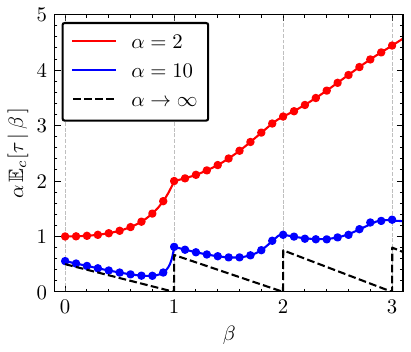}
   \caption{Rescaled conditional mean first-passage time $\alpha \mathbb{E}_c\left[\,\tau\,\vert\,\beta\,\right]$ as a function of the offset $\beta$ for $\alpha=2$ (red) and $\alpha=10$ (blue). Solid lines correspond to the analytical result \eqref{eq:Ec=supercritical_answer}, while circles correspond to direct numerical simulations obtained by generating $10^{3}$ independent trajectories of the Poisson process for each of $30$ equally spaced values of $\beta$ in the range $[0,3]$. The black dashed line shows the asymptotic result for $\alpha\to\infty$ from \eqref{eq:Ec_alpha_large}.}
    \label{fig:Ec(a>1)}
\end{figure}

 
We now express both terms on the right-hand side as infinite series and invert term by term. For the first term, we apply the representation \eqref{eq:critical_sum_identity} and invert using \eqref{eq:L-1(() ) = simple expression} and \eqref{eq:L-1(1/l () ) = incomplete_gamma}. For the second term, we use the series expansion
\begin{equation}\label{eq:series_second_term}
    \frac{1}{ \left( \alpha\lambda-1+e^{-\lambda}\right)^2 }
    = \sum_{j=0}^{\infty}
        \frac{(j+1) \, e^{-j\lambda} }{(1-\alpha\lambda)^{j+2}} ,
\end{equation}
whose inverse Laplace transform is again obtained via residue calculus and reads
\begin{equation}\label{eq:LT_inverse_second_term}
    \mathcal{L}^{-1}_{\lambda\to\beta}
        \left[  \frac{(j+1) \, e^{-j\lambda}}{(1-\alpha\lambda)^{j+2}}  \right]
    = \frac{1}{\alpha^{j+2}}
    \frac{(-1)^j}{j!}\left( \beta-j \right)^{j+1}
    e^{\frac{\beta-j}{\alpha}}
    \theta(\beta-j).
\end{equation}
Combining all contributions, we arrive at the closed-form expression
\begin{multline}\label{eq:Ec=supercritical_answer}
\alpha>1 :\quad 
    \mathbb{E}_c\left[\,\tau\,\vert\,\beta\,\right]
    =\frac{1}{1-S_{\infty}(\beta)}
    \sum_{j=0}^{\lfloor \beta \rfloor}
    \frac{1}{\alpha^{j+1}}
    \frac{(-1)^{j}}{j!} 
    \Bigg\{   
    \frac{ \left(\beta-j\right)^{j} }{2\, (\alpha-1) } 
        e^{\frac{\beta-j}{\alpha}}
    \\  +
    \left(1-\frac{1}{\alpha}\right)
        \left(\beta-j\right)^{j+1} e^{\frac{\beta-j}{\alpha}} 
    - \int_{0}^{\beta-j} \dd y\, y^{j} e^{\frac{y}{\alpha}}
    \Bigg\},
\end{multline}
where the survival probability $S_\infty(\beta)$ is given by \eqref{eq:Sinfty(beta)=sum}. As a consistency check, setting $\beta=0$ reduces the sum to a single term that reproduces the result \eqref{eq:first_two_moments=a>1} obtained for the zero offset case. The comparison between \eqref{eq:Ec=supercritical_answer} and numerical simulations is shown in  Fig.~\ref{fig:Ec(a>1)}.

\par Note that the conditional mean first-passage time is not necessarily a monotonically increasing function of $\beta$. This can be illustrated by examining the large-$\alpha$ limit. Taking the leading order in $1/\alpha$ from \eqref{eq:Ec=supercritical_answer} and \eqref{eq:Sinfty(beta)=sum}, we obtain
\begin{equation}\label{eq:Ec_alpha_large}
    \mathbb{E}_c\left[\,\tau\,\vert\,\beta\,\right]
    \underset{\alpha\to\infty}{=}
    \frac{1}{\alpha}
    \frac{\lfloor \beta + 1 \rfloor}
         {\lfloor \beta + 2 \rfloor}
         \big(\lfloor \beta+1\rfloor -  \beta \big)
    + O\left(\frac{1}{\alpha^2}\right).
\end{equation}
The factor $(\lfloor \beta+1\rfloor - \beta)$ oscillates, demonstrating the non-monotonic behavior. More detailed analysis reveals that this non-monotonicity first appears for $\alpha>2$. Similarly to the subcritical case \eqref{eq:Ec[k]_alpha=0}, one can show (though we omit the details here) that
\begin{equation}\label{eq:Ec[k]_alpha_large}
    \lim_{\alpha\to\infty}  \alpha^k \,
    \mathbb{E}_c\left[\,\tau^k\,\vert\,\beta\,\right] 
    =  \frac{\lfloor \beta + 1 \rfloor}
         {\lfloor \beta + 1 + k \rfloor}
         \big(\lfloor \beta+1\rfloor -  \beta \big)^{k}.
\end{equation}

\par Let us make a brief remark on numerical simulations. In the supercritical regime, a significant fraction of trajectories never cross the boundary, hence the direct sampling approach results in a much lower number of configurations compared to the subcritical case ($10^3$ instead of $10^6$). Moreover, according to \eqref{eq:Sinfty~e^-beta}, the survival probability decays exponentially fast with $\beta$, making it increasingly unlikely for a directly sampled trajectory to cross the boundary at large offsets. In principle, this sampling difficulty could be circumvented using more advanced techniques such as Metropolis algorithms on trajectory space or importance sampling methods \cite{H-25} (for example, introducing an exponential bias via ``local tilt'' \cite{BMR-24} to the maximum distance between the Poisson process trajectory and the moving boundary). However, such an extensive numerical investigation falls beyond the scope of the present work. The analytical results \eqref{eq:Ec=supercritical_answer} and \eqref{eq:Sinfty(beta)=sum} are exact and valid for arbitrary $\beta$, and the simpler direct sampling approach suffices to verify them for moderate values of $\beta$.

\paragraph{Critical behavior of the moments}

\par The final piece in the analysis concerns the behavior of the moments as $\alpha$ approaches the critical value $\alpha\to1$. We expect the moments to diverge, and we now quantify this divergence following the same approach as for the zero offset case. Recall that the generating function of the rescaled moments \eqref{eq:F(R,alpha|beta)=sum_moments} is given by
\begin{equation}\label{eq:F(R,alpha|beta)=sum_moments_general}
    \mathcal{F}(R,\alpha\,\vert\,\beta) 
    =
    \frac{1}{\left|\alpha-1\right|}
    \left[
        \frac{Q\left(R(\alpha-1)^2\,\vert\,\beta\right)}
             {1-S_\infty(\beta)}
        - 1
    \right].
\end{equation}
The divergence of the conditional moments can be extracted by computing the limit $\alpha\to1$ in \eqref{eq:F(R,alpha|beta)=sum_moments_general}. The difference from the zero offset case is that there is no closed-form representation for $Q(\rho\,\vert\,\beta)$, but only for its Laplace transform $\hat{Q}(\rho\,\vert\,\lambda)$. We therefore proceed as follows: first, we expand \eqref{eq:hat(Q)=explicit} in series around $\alpha=1$, keeping terms up to order $(\alpha-1)^2$; second, we use the sum identity \eqref{eq:critical_sum_identity} and invert the Laplace transform term by term; finally, we combine this result with the expression \eqref{eq:Sinfty(beta)=sum} for the survival probability $S_\infty(\beta)$.
This procedure yields
\begin{equation}\label{eq:F_critical_general}
    \lim_{\alpha\to1}
    \mathcal{F}(R,\alpha\,\vert\,\beta) 
    =
    \frac{1}{2}\left(1-\sqrt{1+2R}\right) \sum_{j=0}^{\lfloor \beta \rfloor } 
        \frac{(-1)^{j}}{j!} (\beta-j)^j e^{\beta-j}.
\end{equation}
Expanding in powers of $R$ and comparing with \eqref{eq:F(R,alpha|beta)=sum_moments}, we obtain the divergence of the moments
\begin{equation}
    \mathbb{E}_{c}\left[\,\tau^{k}\,\vert\,\beta\,\right]
    \underset{\alpha\to1}{\sim}
    \frac{(2k)!}{2^{k+1}k! (2k-1)} 
    \left| \frac{1}{\alpha-1} \right|^{2k-1}
    \sum_{j=0}^{\lfloor \beta \rfloor } \frac{(-1)^{j}}{j!} (\beta-j)^j e^{\beta-j}.
\end{equation}
Note that this reduces to \eqref{eq:Ec=divergence_exact} when $\beta=0$, as expected. This result is valid as $\alpha$ approaches the critical value from both above and below.

\section{Conclusion}\label{sec:Conclusion}

In this paper we studied the first-passage properties of the Poisson process with respect to a linear moving boundary. We presented two complementary approaches to the problem: the direct time-domain approach, which relies on path-decomposition techniques and various combinatorial identities and was first presented in \cite{P-59}; and the Laplace-domain approach, an adaptation of the method recently presented in \cite{BM-25}, whose key ingredient is the mapping of the original process onto an effective discrete-time random walk. This mapping allowed us to apply the Pollaczek-Spitzer formula and obtain an explicit expression for the Laplace transform of the first-passage time probability density, thereby rederiving the results found in \cite{BD-57}. Although these results are not new, we presented the derivations in a pedagogical and self-contained manner. By developing both approaches in parallel and illustrating how they complement each other, we hope to make these techniques more accessible to the broader physics community.

\par 
Beyond rederiving existing expressions, we combined the two approaches to obtain several new analytical results. 
Specifically, we derived explicit closed-form expressions for the conditional mean first-passage time for arbitrary offset $\beta$, and obtained a large deviation form for the first-passage time distribution in the subcritical regime ($\alpha<1$), where crossing is certain. We also characterized the large-$\tau$ behavior of the survival probability, showing that it decays exponentially to its asymptotic value with a universal timescale $\xi(\alpha)=(1-\alpha+\alpha\log\alpha)^{-1}$. Finally, we determined the divergence of the conditional moments as the boundary slope approaches its critical value $\alpha\to1$, where the characteristic timescale diverges. With reference to the D/M/1 queueing model discussed in the Introduction, these new results apply immediately and hence extend what is known about the busy period distribution and its associated statistics.

\par 
The problem considered in the present paper is a rare example in which all quantities of interest can be computed explicitly. It would be interesting to investigate to what extent the presented results can be generalized. One natural direction concerns more general boundaries. For this scenario, a number of results based on path-decomposition techniques already exist \cite{T-65,G-66,G-93,Lehmann1998,PSZ-99}, so it is worth asking if the mapping onto an effective random walk can be adapted to such problems and, if so, whether it would yield new explicit expressions as it did for the linear boundary. Another direction would be to consider more general processes (see \cite{Br-25} and references therein). As was shown in \cite{B-25} for the dual process, the Laplace transform technique is quite powerful for extracting the asymptotic behavior of the survival probability for a broad class of processes. Whether this approach can also be used to obtain large deviation functions is an interesting question that remains to be addressed.

\begin{appendix}

\section{Properties of the Lambert W-function}\label{sec:app_LambertW}

\par In this appendix, we summarize the essential properties of the Lambert W-function, also known as the product logarithm, which appears throughout the analysis in this paper. For a more comprehensive treatment, we refer the reader to~\cite{CGHJK-96,CJK-97}.

\par The Lambert W-function is defined as the inverse of the function $f(w) = we^w$. For any complex number $z$, the equation
\begin{equation}\label{eq:app_Lambert_def}
    W(z)\,e^{W(z)} = z
\end{equation}
defines $W(z)$ implicitly.

\begin{figure}[h]
    \includegraphics{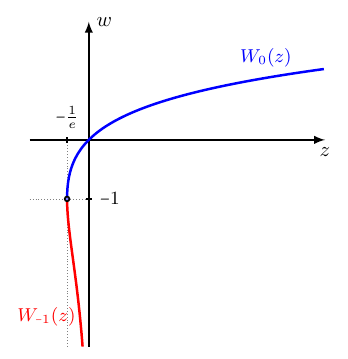}
    \caption{The two real branches of the Lambert W-function: the principal branch $W_0(z)$ (blue) defined for $z \geq -1/e$, and the secondary branch $W_{-1}(z)$ (red) defined for $-1/e \leq z < 0$. Both branches meet at the branch point $(-1/e, -1)$.}
    \label{fig:lambertW01}
\end{figure}

\paragraph{Complex structure} 
The inverse relation~\eqref{eq:app_Lambert_def} is multivalued: for a given $z$, there can be multiple values of $w$ satisfying $we^w = z$. To understand this structure, we first consider the function restricted to real arguments. The function $f(w) = we^w$ for $w \in \mathbb{R}$, shown in Fig.~\ref{fig:lambertW01}, attains a minimum value of $-1/e$ at $w = -1$ and is monotonically increasing on $[-1,\infty)$ and monotonically decreasing on $(-\infty,-1]$. Consequently, for $-1/e \leq z < 0$, there exist two distinct real solutions to $we^w = z$, giving rise to two real branches of the inverse function: the principal branch $W_0(z)$ and the secondary branch $W_{-1}(z)$. For $z \geq 0$, only the principal branch yields a real value.

\par 
The principal branch has range $W_0(z) \geq -1$, while the secondary branch has range $W_{-1}(z) \leq -1$. From this observation, it follows that for any real $z$, the product $ze^z$ can be inverted back to $z$ using the appropriate branch. Specifically, for $z \geq -1$, we have $ze^z \geq -1/e$ with $ze^z$ lying on the increasing portion of $f(w)$, so the principal branch applies:
\begin{equation}
   z \geq -1:\qquad  W_0(ze^z) = z.
\end{equation}
Similarly, for $z \leq -1$, we have $ze^z \geq -1/e$ with $ze^z$ lying on the decreasing portion of $f(w)$, so the secondary branch applies:
\begin{equation}
   z \leq -1:\qquad  W_{-1}(ze^z) = z.
\end{equation}

\par 
In the complex plane, the Lambert W-function has countably many branches, denoted $W_\ell(z)$ where $\ell \in \mathbb{Z}$ is the branch index. The principal branch $W_0(z)$ is defined for all $z \in \mathbb{C}$ with $W_0(0) = 0$, whereas the branches $W_\ell(z)$ with $\ell \neq 0$ are defined for $z \neq 0$ and satisfy $\lim_{z \to 0} W_\ell(z) = -\infty$. The principal branch has a branch point at $z = -1/e$ with a branch cut conventionally taken along the negative real axis from $-1/e$ to $-\infty$. All other branches have a branch point at $z = 0$ with a branch cut along the entire negative real axis.

\par
\begin{figure}[ht]
    \includegraphics{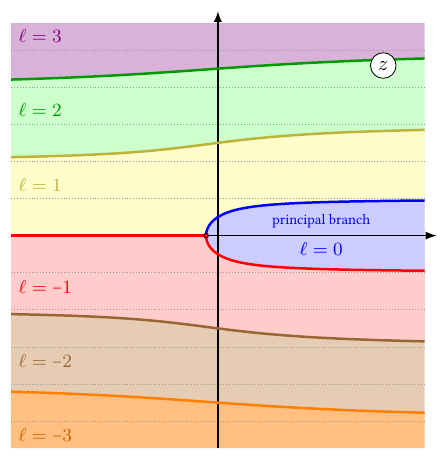}
    \caption{Regions in the complex $z$-plane where the identity $W_\ell(ze^z) = z$ holds for different branch indices $\ell$. Each colored region corresponds to a different branch $W_\ell(z)$. The curves separating adjacent regions are given parametrically by $z = -\eta\cot(\eta) + i\eta$ and are asymptotic to horizontal lines at integer multiples of $\pi i$.}\label{fig:LambertComplexStructure}
\end{figure}

 The inverse relationship generalizes to the complex plane as
\begin{equation}\label{eq:app_Lambert_inverse}
    W_\ell(ze^z) = z,
\end{equation}
where each branch $W_\ell(z)$ satisfies this identity within a specific region of the complex $z$-plane. The structure of these branches and their domains of validity are illustrated in Fig.~\ref{fig:LambertComplexStructure}. The boundaries separating adjacent regions are smooth curves given parametrically by
\begin{equation}
    z(\eta) = -\eta\cot\eta + i\eta,
\end{equation}
where $-\pi<\eta<\pi$ for the principal branch, and $2\ell \pi<\pm\eta< (2\ell+1)\pi$ for other branches.

\paragraph{Derivatives and series expansions.}

Having established the branch structure, we now turn to the 
analytical properties of the Lambert W-function used throughout 
the paper. Differentiating both sides of~\eqref{eq:app_Lambert_def} 
with respect to $z$ yields
\begin{equation}
        \left( e^{W(z)}  + W(z)e^{W(z)}\right) \dv{}{z}\Big[ W(z) \Big]  = 1,
\end{equation}
and hence
\begin{equation}
    \dv{}{z}\Big[ W(z) \Big] = \frac{1}{e^{W(z)}  + W(z)e^{W(z)}}.
\end{equation}
Using~\eqref{eq:app_Lambert_def} to substitute $e^{W(z)} = z/W(z)$, this becomes
\begin{equation}
    \dv{}{z}\Big[ W(z) \Big] = \frac{W(z)}{z(1+W(z))},
\end{equation}
which is the differentiation rule stated in~\eqref{eq:d/dz W_0=rule}.

\par 
Another property we use is the series expansion of the principal branch $W_0(z)$ around $z = 0$. This is obtained via the Lagrange  inversion theorem, which may be formulated as  follows~\cite{G-16}. Let $z = f(w)$ and $z_0 = f(w_0)$ with  $f'(w_0) \neq 0$. Then the inverse function admits a power  series expansion of the form
\begin{equation}\label{eq:app_Lagrange_inversion}
    w = w_0 + \sum_{n=1}^{\infty} \frac{(z - z_0)^n}{n!}
    \lim_{w\to w_0}
    \left\{  \dv{^{n-1}}{w^{n-1}}  \left(\frac{w-w_0}{f(w)-z_0}\right)^{n} \right\}.
\end{equation}
More generally for any analytic function $g(w)$,
\begin{equation}\label{eq:app_Lagrange_inversion_2}
    g(w) = g(w_0) + \sum_{n=1}^{\infty} \frac{(z-z_0)^n}{n!}
    \lim_{w\to w_0}
    \left\{  \dv{^{n-1}}{w^{n-1}} \left[ g'(w)\left(\frac{w-w_0}{f(w)-z_0}\right)^{n} \right] \right\}.
\end{equation}
Setting $f(w) = w e^w$ and $w_0=0$, we obtain the expansion for the principal branch of the Lambert function around the origin in the form:
\begin{equation}\label{eq:app_W0_expansion}
    W_{0}(z) = \sum_{n=1}^{\infty} \frac{z^{n}}{n!} 
    \lim_{w\to 0} \left\{  \dv{^{n-1}}{w^{n-1}} e^{-n w} \right\} 
    = \sum_{n=1}^{\infty} \frac{(-n)^{n-1}}{n!} z^n. 
\end{equation}
Furthermore, setting $g(w) = w^j$ in \eqref{eq:app_Lagrange_inversion_2} gives
\begin{equation}
    \Big[W_0(z)\Big]^j =
    j
    \sum_{n=1}^{\infty} \frac{z^n}{n!}
    \lim_{w\to 0} \left\{  \dv{^{n-1}}{w^{n-1}} \left[ w^{j-1} e^{-nw}
    \right] \right\}.
\end{equation}
Expanding $e^{-nw}$ as a power series and noting that  multiplication by $w^{j-1}$ shifts all powers, the $(n-1)$-th  derivative evaluated at $w = 0$ extracts the coefficient of  $w^{n-1}$. This yields the identity~\eqref{eq:W0^j=expansion}:
\begin{equation}\label{eq:app_W0^j=expansion}
    \Big[W_0(z)\Big]^j 
    = -
    j \sum_{n=j}^{\infty}  \frac{(-n)^{n-j-1}}{(n-j)!}  z^n,
    \quad j\ge 1,
\end{equation}
which in particular implies \eqref{eq:app_W0_expansion}.

\section{Derivation of the Laplace transform}\label{sec:app_Q_Laplace}
\par In this appendix we obtain the explicit representation \eqref{eq:hat(Q)(rho,s|lambda)=explicit} for the triple Laplace transform of the first-passage time of the Poisson process with respect to the linear moving boundary.

\par The starting point is the general result obtained from the Pollaczek-Spitzer formula. According to \eqref{eq:hatQ=PollSpitzer} it reads
\begin{equation}\label{eq:app_hatQ=PollSpitzer}
    \hat{\mathcal{Q}}(\rho,s\,\vert\,\lambda) = 
    \frac{1}{\lambda} - 
    \frac{1 - s\,c(\rho)}{\lambda}
     \phi^{-}(\lambda;\rho,s) \,
    \phi^{+}(0;\rho,s),
\end{equation}
where the functions $\phi^\pm(\lambda;\rho,s)$ are defined as
\begin{equation}\label{eq:app_phi^pm=def}
    \phi^{\pm}(\lambda;\rho,s) = 
    \exp\left[
        -\frac{1}{2\pi} \int_{-\infty}^{\infty}\dd k\, \frac{1}{\lambda\pm\ii k} \log\left[1 - s c(\rho) F(k;\rho)\right]
    \right].
\end{equation}
For the problem at hand, the explicit expressions of $F(k;\rho)$ and $c(\rho)$ are given in \eqref{eq:F(k;rho)=Poisson} and we have
\begin{equation}\label{eq:app_F(k;rho)=Poisson}
    F(k;\rho) = \frac{\rho+1}{\rho+1-\ii\alpha k} \, e^{-\ii k},
    \qquad
    c(\rho) = \frac{1}{\rho+1}.
\end{equation}
Below we provide the explicit computation of the functions $\phi^\pm(\lambda;\rho,s)$, which relies on a standard complex analysis procedure.

\par
First, integration by parts in \eqref{eq:app_phi^pm=def} yields an alternative representation for the functions $\phi^\pm(\lambda;\rho,s)$ in the form
\begin{equation}\label{eq:app_phi^pm=alternative}
    \phi^\pm(\lambda;\rho,s) = 
    \exp\left[ \frac{1}{2\pi\ii} \int_{-\infty}^{\infty} \dd k\, \mathcal{I}^{\pm}(k) \right],
\end{equation}
where
\begin{equation}\label{eq:app_I=def}
    \mathcal{I}^{\pm}(k) \equiv 
    \mp \log\left[k\mp\ii\lambda\right]
    \frac{\pd_k F(k;\rho)}
         {\frac{1}{s c(\rho)} - F(k;\rho)},
\end{equation}
or explicitly
\begin{equation}\label{eq:app_I=explicit}
    \mathcal{I}^{\pm}(k) = 
    \pm \log\left[k\mp\ii\lambda\right]
    \frac{\ii s\,  e^{-\ii k}}
         {\rho+1-\ii \alpha k}
    \frac{\rho+1-\ii \alpha k - \alpha}
         {\rho+1-\ii \alpha k - s e^{-\ii k}}.
\end{equation}
The integrals are then computed by closing the contour of integration in the complex $k$-plane.

\par
\begin{figure}[h]
    \includegraphics{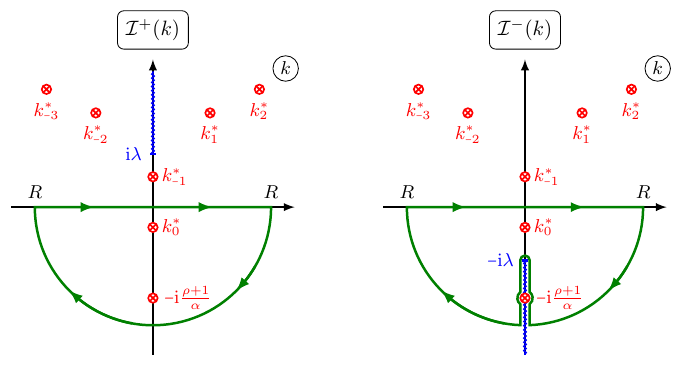}
    \caption{
    Schematic representation of the analytic structure of the integrand $\mathcal{I}^+(k)$ (left) and $\mathcal{I}^{-}(k)$ (right) and the contour closure.  
    The integrals over the semicircles vanish as $R\to\infty$.
    }\label{fig:app_LT_derivation}
\end{figure}

 
We now analyze the analytic structure of the integrands $\mathcal{I}^{\pm}(k)$ in \eqref{eq:app_I=explicit}. There is a branch cut originating from the logarithm; additionally, there are two types of poles: (i) a simple pole corresponding to the zero of the first denominator, $k=-\ii\frac{\rho+1}{\alpha}$; (ii) a set of poles corresponding to the zeros of the second denominator, which occur when
\begin{equation}\label{eq:app_pole_equation}
    \rho+1-\ii \alpha k - s e^{-\ii k} = 0.
\end{equation}
To solve this equation we rewrite it as
\begin{equation}
    \left(\ii k - \frac{\rho+1}{\alpha}\right)
     e^{\left(\ii k - \frac{\rho+1}{\alpha}\right) } = - \frac{s}{\alpha} e^{-\frac{\rho+1}{\alpha}}.
\end{equation}
Setting $u = \left(\ii k - \frac{\rho+1}{\alpha}\right)$, this becomes $u e^u = -\frac{s}{\alpha} e^{-\frac{\rho+1}{\alpha}}$, which is precisely the defining equation for the Lambert function $W(z)$, with solutions $u = W_\ell[z]$ for each branch $\ell \in \mathbb{Z}$. Therefore, the solutions of \eqref{eq:app_pole_equation} are given by
\begin{equation}\label{eq:app_k_poles=}
    k_\ell^*(\rho,s) = -\ii \left(
        \frac{\rho+1}{\alpha}
        + 
        W_{\ell}\left[
            -s \frac{1}{\alpha} e^{-\frac{\rho+1}{\alpha}}
        \right]
    \right),
    \qquad \ell \in \mathbb{Z}.
\end{equation}
The schematic representation of the analytic structure of $\mathcal{I}^{\pm}(k)$ is shown in Fig.~\ref{fig:app_LT_derivation}.

\par Once the analytic structure of the integrands is established, we can compute them by closing the contour in the lower half-plane. The integrals over the semicircles vanish. Note that out of all poles \eqref{eq:app_k_poles=}, only the one at $k_0^*(\rho,s)$ is actually relevant. This simple observation allows us to compute the integral corresponding to $\phi^{+}(\lambda;\rho,s)$ as a sum of two residues:
\begin{equation}
    \frac{1}{2\pi\ii}\int_{-\infty}^{\infty}\dd k\, 
        \mathcal{I}^{+}(k) = 
    - \Res_{k=-\ii\frac{\rho+1}{\alpha}}\left[
        \mathcal{I}^{+}(k)
    \right]
    -\Res_{k=k_0^{*}(\rho,s)}\left[
        \mathcal{I}^{+}(k)
    \right].
\end{equation}
The difference for $\phi^-(\lambda;\rho,s)$ is that apart from the two poles we must take into account the integral along the branch cut of the logarithm:
\begin{equation}
    \frac{1}{2\pi\ii}\int_{-\infty}^{\infty} \dd k\, \mathcal{I}^{-}(k) = 
    - \Res_{k=-\ii\frac{\rho+1}{\alpha}}\left[
        \mathcal{I}^{-}(k)
    \right]
    -\Res_{k=k_0^{*}(\rho,s)}\left[
        \mathcal{I}^{-}(k)
    \right]
    -\frac{1}{2\pi\ii}
    \int_{\text{b.c.}} \dd k\, \mathcal{I}^{-}(k),
\end{equation}
where ``b.c.'' denotes the branch cut of the logarithm, and the integral is given by
\begin{equation}
    -\frac{1}{2\pi \ii}\int_{\text{b.c.}}
    \dd k\, \mathcal{I}^{-}(k)
    = -
    \int_{-\ii\infty}^{-\ii\lambda}
        \dd k \, 
        \frac{\pd_k F(k;\rho)}
         {\frac{1}{s c(\rho)} - F(k;\rho)}.
\end{equation}
Recognizing this as a total derivative and using $F(-\ii\infty;\rho)=0$, we immediately conclude that
\begin{equation}
    -\frac{1}{2\pi \ii}\int_{\text{b.c.}}
    \dd k\, \mathcal{I}^{-}(k)
    =
    - \log\left[1 - s c(\rho) F(-\ii\lambda;\rho)\right].
\end{equation}
The only thing left is to compute the residues. A straightforward calculation shows that
\begin{equation}
     \Res_{k=-\ii\frac{\rho+1}{\alpha}}\left[
        \mathcal{I}^{\pm}(k)
    \right]
    = \mp \log\left[
        -\ii \left(\frac{\rho+1}{\alpha}\pm\lambda\right)
    \right],
\end{equation}
and
\begin{equation}
    \Res_{k=k_0^{*}(\rho,s)}\left[
        \mathcal{I}^{\pm}(k)
    \right]
    = \pm\log\left[k_0^{*}(\rho,s)\mp\ii \lambda\right].
\end{equation}
Combining everything together, we obtain the explicit 
representations for $\phi^\pm(\lambda;\rho,s)$:
\begin{equation}\label{eq:phi^+=explicit}
    \phi^{+}(\lambda;\rho,s)
    =
    \frac{\frac{\rho+1}{\alpha} +\lambda}{\ii k_0^*(\rho,s)+\lambda},
\end{equation}
and
\begin{equation}\label{eq:phi^-=explicit}
    \phi^{-}(\lambda;\rho,s)
    =
    \frac{\ii k_0^*(\rho,s) - \lambda}{\frac{\rho+1}{\alpha}-\lambda} 
    \frac{1}{1-sc(\rho)F\left(-\ii\lambda;\rho\right)}.
\end{equation}
Substituting \eqref{eq:phi^+=explicit} and \eqref{eq:phi^-=explicit} into \eqref{eq:app_hatQ=PollSpitzer} and using the explicit form of $k_0^{*}(\rho,s)$ as in \eqref{eq:app_k_poles=}, after algebraic manipulations, we obtain 
\begin{equation}\label{eq:app_hat(Q)(rho,s|lambda)=explicit}
    \hat{\mathcal{Q}}(\rho,s\,\vert\,\lambda) =
    \frac{1}{\lambda} 
    - \frac{1}{\lambda}\frac{\rho+1-s}{\rho+1-\alpha\lambda - s \, e^{-\lambda}}
    \left(1 - \frac{\alpha\lambda}{\rho+1 + \alpha W_0\left[-s \frac{1}{\alpha}e^{-\frac{\rho+1}{\alpha}}\right]}\right),
\end{equation}
which is exactly the result stated in \eqref{eq:hat(Q)(rho,s|lambda)=explicit}.

\section{Asymptotics of the survival probability. Time-domain approach}\label{sec:app_Sinfty}

\par 
In this appendix we analyze the $\tau\to\infty$ behavior of the survival probability $S(\tau\,\vert\,\beta)$ using the representation \eqref{eq:S(t,beta)=result_time_domain} obtained via the direct time-domain approach,
\begin{equation}
\label{eq:app_S(t,beta)=result_time_domain}
    S(\tau\,\vert\,\beta) = \sum_{n=0}^{\lfloor\alpha \tau+\beta
    \rfloor} 
        \frac{(\alpha \tau + \beta - n)}{\alpha^n} 
        \sum_{j=0}^{\lfloor\beta\rfloor} 
        \frac{(j-\beta)^j(\alpha \tau + \beta - j)^{n-j-1}}
             {j!(n-j)!}e^{-\tau}.
\end{equation}
The key idea is to recognize in the summation above the upper incomplete gamma function, a generalization of the usual gamma function, defined as
\begin{equation}\label{eq:app_upper_Gamma=def}
    \Gamma(m,z) \equiv \int_{z}^{\infty} \dd u\, u^{m-1} e^{-u}.
\end{equation}
Integration by parts in \eqref{eq:app_upper_Gamma=def} yields an alternative representation for integer $m$:
\begin{equation}\label{eq:app_upper_Gamma=sum}
    \Gamma(m,z) = (m-1)! \sum_{k=0}^{m-1} \sigma(k,z),\qquad
    \sigma(k,z) \equiv \frac{z^{k}}{k!}e^{-z}.
\end{equation}
Recognizing the sum \eqref{eq:app_upper_Gamma=sum} in \eqref{eq:app_S(t,beta)=result_time_domain} and replacing it by the integral representation \eqref{eq:app_upper_Gamma=def} allows us to perform the asymptotic analysis $\tau\to\infty$ in \eqref{eq:app_S(t,beta)=result_time_domain}.

\par 
We emphasize that, similarly to the discussion presented in Section~\ref{subsec:equivalence}, where we established the equivalence between the time-domain and Laplace-domain approaches by recognizing series representations of the Lambert function in the expression for $S(\tau\,\vert\,0)$, the derivation presented below is motivated by the Laplace transform approach. In some sense, although not stated explicitly, the incomplete gamma function appears when computing the mean first-passage time for general offset $\beta$ (recall \eqref{eq:L-1(1/l () ) = incomplete_gamma}). This partially motivates the search for it in the representation \eqref{eq:app_S(t,beta)=result_time_domain}, once again demonstrating the complementary nature of the two approaches.

\par We now demonstrate how the incomplete gamma function emerges from \eqref{eq:app_S(t,beta)=result_time_domain}. By changing the order of summation and performing algebraic manipulations, we rewrite it as
\begin{equation}
    S(\tau\,\vert\,\beta) = 
    \sum_{j=0}^{\lfloor\beta\rfloor}
    \frac{(j-\beta)^{j}}{\alpha^j j!}
    e^{\frac{\beta-j}{\alpha}}
    \sum_{n=0}^{\lfloor\alpha \tau+\beta\rfloor} 
        \frac{e^{-\tau - \frac{\beta-j}{\alpha}}}{(n-j)!}
        \left(\tau + \frac{\beta - n}{\alpha}\right)
        \left(\tau + \frac{\beta-j}{\alpha}\right)^{n-j-1}.
\end{equation}
Note that due to the factor $(n-j)!$, all terms with $j>n$ vanish. Then introducing an auxiliary quantity
\begin{equation}\label{eq:app_tau_j=def}
    \tau_j \equiv \tau + \frac{\beta-j}{\alpha}
\end{equation}
and changing the summation index $n\mapsto k = n-j$, we arrive at
\begin{equation}\label{eq:app_S(tau|beta)=sum_tau j}
    S(\tau\,\vert\,\beta) = 
    \sum_{j=0}^{\lfloor\beta\rfloor}
    \frac{(j-\beta)^{j}}{\alpha^j j!}
    e^{\frac{\beta-j}{\alpha}}
    \sum_{k=0}^{\lfloor\alpha \tau_j\rfloor} 
        \frac{e^{-\tau_j}}{k!}
        \left(\tau_j - \frac{k}{\alpha}\right)
        \tau_j^{k-1}. 
\end{equation}  
Comparing now the latter sum with \eqref{eq:app_upper_Gamma=sum}, we recognize the expression for the incomplete function, hence 
\begin{equation}\label{eq:app_S(tau|beta)=sum_Gamma}
     S(\tau\,\vert\,\beta)
     = 
    \sum_{j=0}^{\lfloor\beta\rfloor}    
    \frac{(j-\beta)^{j}}{\alpha^j j!}
    e^{\frac{\beta-j}{\alpha}}
   \left\{
       \left(1-\frac{1}{\alpha}\right) 
       \frac{\Gamma\left(\lfloor \alpha \tau_j\rfloor, \tau_j\right)}
            {(\lfloor \alpha \tau_j \rfloor-1)!}
        + 
        \sigma\left(\lfloor \alpha\tau_j\rfloor, \tau_j\right)
   \right\}.
\end{equation} 
To extract the asymptotic behavior of \eqref{eq:app_S(tau|beta)=sum_Gamma}, we evaluate the survival probability at the discrete set of times $\tau = T_n$ defined in \eqref{eq:Tn=def}, which represent the earliest times at which the $n$-th jump can occur without crossing the boundary. This choice significantly simplifies the technical derivation.

\par 
Since the survival probability is monotonically decreasing, we have
\begin{equation}
    T_{n} < \tau < T_{n+1}: \qquad 
    S\left(T_{n+1}\,\vert\,\beta\right) 
    < S\left(\tau\,\vert\,\beta\right) 
    < S\left(T_{n}\,\vert\,\beta\right).
\end{equation}
Therefore, the $n \to \infty$ behavior at these discrete times captures the $\tau \to \infty$ asymptotics. For sufficiently large $n$, we have $T_n = \frac{n-\beta}{\alpha}$, hence $\tau_j=\frac{n-j}{\alpha}$ and \eqref{eq:app_S(tau|beta)=sum_Gamma} becomes
\begin{equation}\label{eq:app_S(n|beta)=sum_Gamma}
    S\left(\frac{n-\beta}{\alpha}\,\Big\vert\,\beta\right) = 
    \sum_{j=0}^{\lfloor\beta\rfloor}    
    \frac{(j-\beta)^{j}}{\alpha^j j!}
    e^{\frac{\beta-j}{\alpha}}
   \left\{
       \left(1-\frac{1}{\alpha}\right) 
       \frac{\Gamma\left(n-j, \frac{n-j}{\alpha}\right)}
            {(n-j-1)!}
        + 
        \sigma\left(n-j, \frac{n-j}{\alpha}\right)
   \right\}.
\end{equation}
We now construct the $n\to\infty$ asymptotic expansion using two known results. First, the Stirling asymptotic expansion for the gamma function:
\begin{equation}\label{eq:app_Gamma(a)=Stirling}
    \Gamma(a) \underset{a\to\infty}{=} e^{-a} a^{a}\sqrt{\frac{2\pi}{a}} \sum_{k=0}^{\infty} 
    (-1)^k \frac{\gamma_k}{a^k},
\end{equation}
where the coefficients $\gamma_k$ are known explicitly. The first three of them are
\begin{equation}
    \gamma_0=1,\qquad 
    \gamma_1 = -\frac{1}{12},
    \qquad 
    \gamma_2 = \frac{1}{288}.
\end{equation}
Second, for large $a$, the upper incomplete gamma function admits the uniform asymptotic expansion~\cite{T-79}
\begin{equation}\label{eq:app_Gamma-upper-uniform}
  \frac{\Gamma(a,z)}{\Gamma(a)} \underset{a\to\infty}{=} 
  \frac{1}{\sqrt{2\pi}} \int_{\eta\sqrt{a}}^\infty \dd u\, e^{-\frac{u^2}{2}} 
    + \frac{e^{-\frac{1}{2}a\eta^{2}}}{\sqrt{2\pi a}}\,
      \sum_{k=0}^{\infty} \frac{c_k(\eta)}{a^{k}}
  ,
\end{equation}
where
\begin{equation}\label{eq:app_eta=def}
  \eta = \mathrm{sign}\left(\frac{z}{a}-1\right)
    \sqrt{2\left(\frac{z}{a} - 1 - \ln \frac{z}{a} \right)},
\end{equation}
and the coefficients $c_k(\eta)$ are polynomials in $\eta$ generated by the recurrence relation (with $\gamma_k$ from \eqref{eq:app_Gamma(a)=Stirling})
\begin{equation}\label{eq:app_ck=recurrence}
    c_k(\eta) = \frac{\gamma_k}{\frac{z}{a}-1} 
        + \frac{1}{\eta} \dv{}{\eta} \Big[ c_{k-1}(\eta)\Big],
    \qquad 
    c_0(\eta) = \frac{1}{\frac{z}{a} - 1} - \frac{1}{\eta}.  
\end{equation}
For example, the first coefficient $c_1(\eta)$ reads
\begin{equation}\label{eq:app_c1=result}
  c_{1}(\eta) = \frac{1}{\eta^3} 
    - \frac{1}{\left(\frac{z}{a} - 1\right)^3} 
    - \frac{1}{\left(\frac{z}{a} - 1\right)^2} - \frac{1}{12\left(\frac{z}{a}-1\right)}.
\end{equation}
We stress that $\eta$ and $\frac{z}{a}$ in the recurrence relation \eqref{eq:app_ck=recurrence} are not independent and are related via \eqref{eq:app_eta=def}.

\par 
We now apply these asymptotic expansions to the terms appearing in \eqref{eq:app_S(n|beta)=sum_Gamma}. A crucial observation is that the expansion \eqref{eq:app_Gamma-upper-uniform} is uniform with respect to $\frac{z}{a}$, which ensures the validity of the asymptotic analysis for all terms in the sum. Therefore, setting $a = m$ and $z = m/\alpha$ in \eqref{eq:app_Gamma-upper-uniform} and retaining terms up to order $O(m^{-2})$, we obtain
\begin{multline}\label{eq:app_Gamma(m,m/a)=asympt}
    \frac{\Gamma\left(m, \frac{m}{\alpha}\right)}{(m-1)!}
    \underset{m\to\infty}{\sim}
    \frac{1}{\sqrt{2\pi}}\int_{\mu \sqrt{m}}^{\infty} \dd y\, e^{-\frac{y^2}{2}}
    + 
    \frac{e^{- m \frac{\mu^2}{2} }}{\sqrt{2\pi m}}
    \left( \frac{\alpha}{1-\alpha} - \frac{1}{\mu} \right)
    \\
    +
    \frac{1}{m} \frac{e^{- \frac{m \mu^2}{2}}}{\sqrt{2\pi m}}
    \left[\frac{1}{\mu^3} - \frac{\alpha^2}{(1-\alpha)^3} - \frac{\alpha}{12(1-\alpha)}\right],
\end{multline}
where
\begin{equation}
    \mu = \mathrm{sign}\left(1-\alpha\right)\sqrt{2\left(\frac{1-\alpha}{\alpha} + \log\alpha\right)}.
\end{equation}
Similarly, applying the Stirling expansion \eqref{eq:app_Gamma(a)=Stirling}, we find
\begin{equation}\label{eq:app_sigma(m,m/a)=asympt}
    \sigma\left(m, \frac{m}{\alpha}\right) 
    = \frac{1}{m!} \left(\frac{m}{\alpha}\right)^{m} e^{- \frac{m}{\alpha}}
    \underset{m\to\infty}{\sim} 
    \frac{e^{-m\frac{\mu^2}{2}}}{\sqrt{2\pi m}}\left( 1- \frac{1}{12m} \right).
\end{equation}
The final piece is the asymptotic expansions for the error function
\begin{align}
    &\frac{1}{\sqrt{2\pi}}\int_{z}^{\infty} \dd u\, e^{-\frac{u^2}{2}}
    \underset{z\to\infty}{\sim} 
    \frac{1}{\left|z\right|\sqrt{2\pi}} e^{-\frac{z^2}{2}} \left(1 - \frac{1}{z^2}\right),\\
    &\frac{1}{\sqrt{2\pi}}\int_{z}^{\infty} \dd u\, e^{-\frac{u^2}{2}}
    \underset{z\to-\infty}{\sim} 
    1 - \frac{1}{\left|z\right| \sqrt{2\pi}} e^{-\frac{z^2}{2}} \left(1 - \frac{1}{z^2}\right).
\end{align}

\par With the asymptotic expansions \eqref{eq:app_Gamma(m,m/a)=asympt} and \eqref{eq:app_sigma(m,m/a)=asympt} now established, the remainder of the derivation reduces to straightforward algebraic manipulations. We first observe that the dominant contribution comes from the integral term in \eqref{eq:app_Gamma(m,m/a)=asympt}, which behaves differently depending on whether $\alpha$ is greater or less than unity. Combining the expansions, we deduce that
\begin{equation}
    \lim_{m\to\infty} \left\{
       \left(1-\frac{1}{\alpha}\right) 
       \frac{\Gamma\left(m, \frac{m}{\alpha}\right)}
            {(m-1)!}
        + 
        \sigma\left(m, \frac{m}{\alpha}\right)
   \right\}
   = \begin{cases}
       0, \quad \alpha\le 1\\
       1 - \frac{1}{\alpha}, \quad \alpha>1.
   \end{cases}
\end{equation}
This immediately yields the survival probability at infinite
time. In the subcritical regime,
\begin{equation}
    \alpha\le 1:\qquad 
    S_{\infty}(\beta) =  \lim_{n\to\infty} S\left(\frac{n-\beta}{\alpha}\,\Big\vert\,\beta\right)
    = 0,
\end{equation}
whereas in the supercritical regime,
\begin{equation}
    \alpha>1:\qquad 
    S_{\infty}(\beta) = \lim_{n\to\infty} S\left(\frac{n-\beta}{\alpha}\,\Big\vert\,\beta\right)
    = 
    \left(1-\frac{1}{\alpha}\right)
    \sum_{j=0}^{\lfloor\beta\rfloor}    
    \frac{(j-\beta)^{j}}{\alpha^j j!}
    e^{\frac{\beta-j}{\alpha}}.
\end{equation}
These expressions coincide with the result \eqref{eq:Sinfty(beta)=sum} obtained from the Laplace-domain approach, thereby confirming the equivalence of both methods.

\par We now turn to the asymptotic behavior of $S(\tau|\beta)$ as $\tau \to \infty$. Using the higher-order terms in \eqref{eq:app_Gamma(m,m/a)=asympt} and \eqref{eq:app_sigma(m,m/a)=asympt}, together with the asymptotic expansions of the complementary error function,
we obtain the leading-order behavior
\begin{multline}\label{eq:app_S-Sinf=asmympt}
    S\left(\frac{n-\beta}{\alpha}\,\Big\vert\,\beta\right) - S_{\infty}(\beta) 
    \underset{n\to\infty}{\sim}
    \exp\left[ - \frac{n}{\alpha \; \xi(\alpha)}  \right]
    \\ \times 
    \frac{1}{n^{\frac{3}{2}}} \; \frac{1}{\sqrt{2\pi} }
    \frac{\alpha}{(\alpha-1)^2}
    \sum_{j=0}^{\lfloor\beta\rfloor}    
    \frac{(j-\beta)^{j}}{\alpha^j j!}
    e^{\frac{\beta-j}{\alpha}},
\end{multline}
where the characteristic time scale is
\begin{equation}
    \xi(\alpha) = \frac{2}{\alpha \mu^2} = \frac{1}{1-\alpha + \alpha\log\alpha}.
\end{equation}
The exponential decay rate in \eqref{eq:app_S-Sinf=asmympt} is in perfect agreement with \eqref{eq:S_universal_decay}, obtained from the Laplace-domain approach, once again confirming the consistency of both methods.

\par 
As a final remark, we note that a similar analysis can be applied to study the large-$\beta$ behavior of $S_{\infty}(\beta)$. By evaluating the survival probability at the discrete points $\beta = n$ and applying the Stirling approximation \eqref{eq:app_Gamma(a)=Stirling}, one can show that
\begin{equation}
    1-S_{\infty}(n) 
    \underset{n\to\infty}{\sim} \left(1-\frac{1}{\alpha}\right)
    \frac{e^{n}}{\alpha^n \sqrt{2\pi}}
    \sum_{k=0}^{\infty} \frac{k^{k+n}}{(k+n)^{k+n+\frac{1}{2}}} e^{-k \frac{\mu^2}{2}}.
\end{equation}
This sum can be approximated by an integral 
\begin{equation}
    1-S_{\infty}(n) 
    \underset{n\to\infty}{\sim}
    \left(1-\frac{1}{\alpha}\right) \frac{e^{n}}{\alpha^n}\sqrt{\frac{n}{2\pi}}
    \int_{0}^{\infty} \dd u\, \left(\frac{u}{u+1}\right)^{un + n}
    \frac{e^{- u n \frac{\mu^2}{2}}}{\sqrt{u+1}} .
\end{equation}
Evaluating this integral in the saddle-point approximation recovers the exponential decay \eqref{eq:Sinfty~e^-beta} obtained earlier from the Laplace-domain approach. We omit the technical details, as this calculation does not introduce new insights beyond confirming the consistency of the two approaches.

\end{appendix}

\subsection*{Acknowledgements}
INB and SNM acknowledge support from ANR Grant No. ANR-23-CE30-0020-01 EDIPS.

\subsection*{Data availability statement} 
Data sharing is not applicable to this article as no datasets were generated or analyzed during the current study.

\subsection*{Competing interests}
The authors have no relevant financial or non-financial interests to disclose.

\sloppy
\printbibliography

\end{document}